\documentclass[11pt]{article}

       \newcommand{\Cc}{ {\mathcal{C}} }
       \newcommand{\Dc}{ {\mathcal{D}} }

       \newcommand{\Hc}{ {\mathcal{H}} }

       \newcommand{\Lc}{ {\mathcal{L}} }

  \newcommand{\nn}{\nonumber}
  \newcommand{\bsv}{\boldsymbol{v}}
    \newcommand{\bsJ}{\boldsymbol{J}}
  \newcommand{\bsx}{\boldsymbol{x}}
  
  \newcommand{\bnb}{\boldsymbol{\nabla}}

  \newcommand{\bsE}{\boldsymbol{E}}
  
  \newcommand{\bsxi}{\boldsymbol{\xi}}

  \newcommand{\bsP}{\boldsymbol{P}}
  
  \newcommand{\bsM}{\boldsymbol{M}}

  \newcommand{\nb}{\nabla}

  \newcommand{\bszeta}{\boldsymbol{\zeta}}
  \newcommand{\bseta}{\boldsymbol{\eta}}

\usepackage{graphicx}
\usepackage{epstopdf, epsfig}
\usepackage[colorlinks]{hyperref}
\usepackage{amsfonts}
\usepackage{amssymb}
\usepackage{mathrsfs}
\usepackage[T1]{fontenc}
\usepackage{mathtools}
\usepackage{amsfonts}
\usepackage{dcolumn}
\usepackage{bm}
\usepackage[export]{adjustbox}
\usepackage{wrapfig}
\usepackage{lipsum}
\usepackage{authblk}
\usepackage[font=small,labelfont=bf]{caption}
\allowdisplaybreaks
\usepackage[titletoc,toc,title]{appendix}	

\usepackage[style=ieee, citestyle=numeric-comp, backend=biber]{biblatex}

\usepackage[utf8]{inputenc}
\usepackage{newunicodechar}
\newunicodechar{ʼ}{'}

\usepackage{blindtext}
\usepackage[a4paper,
            bindingoffset=0.2in,
            left=1in,
            right=1in,
            top=1in,
            bottom=1in,
            footskip=.25in]{geometry}

\usepackage{hyperref}
\hypersetup{
    colorlinks=true,
    linkcolor=red,
    filecolor=magenta,      
    urlcolor=cyan,
    citecolor = cyan,
    }

\title{Imposing quasineutrality on electrostatic plasmas\\ via the Dirac theory of constraints} 
\author[1,2]{D. A. Kaltsas\thanks{kaltsas.d.a@gmail.com}}
\author[3]{J. W. Burby}
\author[3]{P. J. Morrison}
\author[4]{\\E. Tassi}
\author[1]{G. N. Throumoulopoulos}
\affil[1]{Department of Physics, University of Ioannina, Ioannina, GR 451 10, Greece}
\affil[2]{Department of Informatics, Democritus University of Thrace, Kavala, GR 654 04, Greece}
\affil[3]{Department of Physics and Institute for Fusion Studies,
The University of Texas at Austin, Austin, TX 78712, USA}
\affil[4]{Université Côte d’Azur, CNRS, Observatoire de la Côte d’Azur, Laboratoire J. L. Lagrange,
Boulevard de l’Observatoire, CS 34229, 06304 Nice Cedex 4, France}
\date{}
 
\begin{document}
\maketitle
\begin{abstract}
We present a method for imposing quasineutrality and, more generally, charge density conservation in the Vlasov-Poisson (VP) and Vlasov-Ampère (VA) systems, which describe electrostatic plasma dynamics, by applying the Dirac theory of constraints. Leveraging the Hamiltonian field formulations of the VP and VA models, we construct generalized Dirac brackets using the Dirac algorithm. The resulting constrained systems enforce charge density conservation, and consequently quasineutrality, given that the initial charge density is zero, through new advection terms in the Vlasov equations involving generalized-force terms, while the electric field is eliminated from the constrained Vlasov dynamics. To verify charge density conservation we conduct one-dimensional numerical experiments using a semi-Lagrangian method, demonstrating that the enforcement of the quasineutrality constraint significantly modifies the dynamics. This approach enables us to identify the forces required to enforce quasineutrality, offering a systematic way to assess the validity of the quasineutral approximation across different kinetic scales.
\end{abstract}

\section{Introduction}
\label{sec_I}

The kinetic description of laboratory and space plasmas is based on a set of Boltzmann-like equations that determine the evolution of the distribution functions for each species in phase space (i.e., position and velocity space), coupled with Maxwell’s equations. In the absence of collisions, these equations reduce to the Vlasov equations, which describe collisionless dynamics and conserve phase-space volume.

For electrostatic phenomena,  where magnetic fields can be neglected, the Vlasov-Maxwell (VM) system reduces to a set of Vlasov equations (typically one for electrons and one for each ion species) coupled with an equation determining the self-consistent electric field. This field can be obtained either from Gauss’s law for electrostatics, leading to the Vlasov-Poisson (VP) system, or from Ampère’s law with the displacement current retained and the magnetic field neglected, resulting in the so-called Vlasov-Amp\`ere (VA) system. The VA system has the advantage that it does not require the imposition of boundary conditions for the electric field and it avoids the need to solve an elliptic equation; instead, the electric field is advanced forward in time using the current distribution at each time step. On the other hand, the VA formulation becomes inconsistent in more than one spatial dimension unless the condition $\bnb \times \bsE = 0$ is satisfied throughout the entire time evolution, since any violation would lead to the self-consistent generation of a magnetic field. However, this condition does not necessarily hold in more than one dimension within the VA framework, whereas this issue does not arise in the VP formulation, where the electric field is defined as $\bsE = -\bnb \phi$ and is therefore inherently irrotational. Therefore, although we use the same 3D notation for both the VP and VA models to maintain notational uniformity, we emphasize that the VA model is physically and mathematically consistent only in 1D. In higher dimensions, the VA system should be replaced by the complete Vlasov--Maxwell system, which will be the topic of a subsequent study.
	
While ordinary magnetohydrodynamics (MHD) and its generalizations, such as Hall MHD and extended MHD, which incorporate ion-drift and electron-inertia effects, respectively, are based on the assumption of quasineutrality (i.e. equal densities of positive and negative charges at every point in space and time, $n_i=n_e$), this is not generally the case for kinetic models such as the VM, VP, and VA systems. These kinetic models are typically employed to study phenomena occurring at small spatial and temporal scales, where deviations from quasineutrality can become significant. On  the  other hand, kinetic effects can also play an important role even in the quasineutral (QN) regime. For this reason, several studies have explored QN limits of Vlasov‑based models, such as the VP, VA, or VM dynamics, aiming to retain the kinetic description in QN plasmas. One successful approach is based on  asymptotic-preserving methods for the QN limit, introduced initially for fluid models \cite{Crispel2005a,Crispel2005b,Crispel2007} and subsequently for the kinetic VP model in \cite{Degond2006,Belaouar2009,Degond2010}. In this approach the self-consistent calculation of the electric field is possible even in the QN case through a reformulation of the Poisson equation, made possible by considering fluid moments of the Vlasov equation.  In \cite{Taitano2024} the authors consider the QN limit of a reformulated Vlasov-Amp\`ere system closed with fluid moments of the Vlasov equation and separating fast and slow dynamics, while in \cite{Blaustein2025}, another asymptotic preserving scheme for the QN limit of the VP system is presented. In the latter work, the VP system is reformulated as a hyperbolic problem applying spectral expansion of Hermite functions in velocity space and constructing an appropriate structure-preserving scheme.

In addition, it is of interest to estimate the length scales at which quasineutrality becomes a valid assumption, by quantifying the forces required to enforce it. The magnitude of these forces can serve as a diagnostic for assessing the validity of the QN approximation across different spatial and temporal scales. The present work focuses on identifying these forces, which are necessary to impose quasineutrality ($n_i=n_e$) to VP and VA dynamics using the Dirac theory of constraints \cite{Dirac1950,Dirac1958,Sundermeyer1982,Morrison2020} and exploiting the generalized Hamiltonian description of VP and VA models, which can be deduced from the most general VM Hamiltonian formulation \cite{Morrison1980b,Morrison1982,Marsden1982}.  One main advantage of using Dirac's theory of constraints, compared to other approaches, is that it yields a constrained system which automatically preserves the Hamiltonian character of the parent unconstrained system, while the constraints become invariant quantities of the new Hamiltonian system, so their preservation is guaranteed.

This article is the second in a two-paper sequence investigating the Hamiltonian structure of QN plasma models. Assuming two space dimensions and frozen background ions, the first paper \cite{Burby2025} showed that the quasineutrality constraint studied here, namely local charge neutrality together with current incompressibility, arises naturally as the first term in the QN slow manifold when the Debye length is much less than the field scale length. It then deduced the Poisson bracket \emph{on} the constraint manifold using the theory of Poisson-Dirac submanifolds. In this paper we consider a general unmagnetized electron-ion plasma in any number of space dimensions when working with the VP model. We also consider both the Vlasov-Poisson and Vlasov-Amp\'ere electrostatic kinetic models. By showing that the QN constraint manifold is amenable to Dirac constraint theory, we provide a sharper picture of the QN Hamiltonian structure than obtained in \cite{Burby2025}. In particular, we find Poisson brackets in an open neighborhood of the constraint manifold that render the constraint a Casimir.  Moreover, while we study the same constraint as in \cite{Burby2025}, here we do not use the quasineutrality ordering parameter to simplify the form of the Hamiltonian. These two studies advance the understanding of QN plasma dynamics in complementary directions, with both ultimately aiming at the full Vlasov-Maxwell system \cite{Burby2015} which will be the subject of a forthcoming paper. 

The Poisson-Dirac constraint method (PDCM) \cite{Pinto2025} and the Dirac constraint method (DCM) give closely related results whenever the Dirac constraint method works, as it does in this case.  Specifically, the PDCM is a generalization of the standard DCM, designed to handle cases where the matrix of Poisson brackets between constraints (the constraint matrix) is not invertible, within a coordinate-free framework. A related approach was developed in \cite{Chandre2015}, where it was proposed to replace the inverse of the constraint matrix with its Moore–Penrose pseudoinverse. This enables the construction of a well-defined Poisson bracket even when first- and second-class constraints are intertwined, but it remains formulated in a coordinate-dependent setting. The PDCM extends these ideas by reformulating the reduction in the language of Poisson–Dirac submanifolds and thus provides a geometric, coordinate-free framework that naturally encompasses both regular and degenerate cases. Conversely, in DCM the constraint manifold corresponds to a special class of Poisson–Dirac manifolds known as Poisson transversals. This stronger geometric condition ensures that the Dirac bracket defines a Poisson structure not only on the constraint manifold itself but also in a neighborhood surrounding it. As a result, the constraint functions become Casimirs of the extended bracket, guaranteeing their preservation under the Hamiltonian flow. When the constraint manifold is merely Poisson–Dirac (and not transversal), this neighborhood extension may fail, and the Poisson structure is then generally restricted to the constraint manifold itself. Technically, to apply the DCM we need to invert the constraint matrix, whereas for the PDCM less restrictive conditions apply, specifically, those required for the existence of the pseudoinverse introduced in \cite{Chandre2015}.

In conclusion: (1) Both methods produce equivalent brackets \emph{on} the constraint manifold. (2) The Poisson-Dirac constraint method does not always give a bracket in a neighborhood of the constraint. However, when it does give such a bracket, the requirements are less restrictive than those needed for the Dirac constraint method to work, but these requirements were not assessed in \cite{Burby2025}. (3) When both methods apply, the brackets obtained in a neighborhood of the constraint manifold will agree as long as the same level set function is used in each method. (Here, level set function refers to the function $C$ such that $C = 0$ defines the constraint).
This is why showing that the Dirac constraint method applies, gives a sharper picture than merely showing the Poisson-Dirac method applies. In geometric terms, showing the Poisson-Dirac method works is the same as showing the constraint manifold is a Poisson-Dirac submanifold, while showing the Dirac method works is the same as showing the constraint manifold is a so-called Poisson transversal. Every Poisson transversal is a Poisson-Dirac submanifold, but the converse is not true.

The Dirac method has been employed previously in continuum models for the imposition of incompressibility constraint, $\bnb\cdot \textbf{v}=0$, within the Eulerian variable
description of ideal hydrodynamics \cite{Nguyen1999,Nguyen2001} and ideal plasma models \cite{Morrison2009,Chandre2012,Chandre2013,Chandre2014,Morrison2020}, exploiting their generalized Hamiltonian formulations. Specifically, the work \cite{Chandre2013}, among other examples, also considered the VM and VP cases emphasizing on the role of projector operators in the context of Dirac constrained dynamics. In that work, quasineutrality as a Dirac constraint was also examined, deriving the associated Dirac projector. However, that paper did not delve deeper in more details or derive a system of constrained equations through the Dirac bracket, which is done in the present work, where the constrained system is constructed via the Dirac algorithm and numerical examples are presented. 

This manuscript is organized as follows: In Section \ref{sec_II}, we present the noncanonical Hamiltonian formulations of the VP and VA models. In Section \ref{sec_III}, we employ the Dirac algorithm to construct generalized Poisson brackets called Dirac brackets, which incorporate the imposed constraints as Casimir invariants. Using these generalized brackets and the Hamiltonians of the VP and VA models, we derive the Dirac-constrained dynamics, which preserve the charge density distribution. So, if the initial charge density is zero, it remains zero throughout the evolution. In Section \ref{sec_IV}, we describe a semi-Lagrangian numerical method	 and present simulation results for the two-stream instability, comparing the constrained QN dynamics to the fully electrostatic VP dynamics with a self-consistent electric field. Finally, in Section \ref{sec_V}, we summarize the results and discuss directions for future work.

\section{Hamiltonian formulation of the Vlasov-Poisson system and the Dirac algorithm}
\label{sec_II}

\subsection{The Vlasov-Poisson system}

A fully ionized, collisionless, electrostatic electron-ion plasma, where magnetic field contributions can be neglected, can be described kinetically by the Vlasov-Poisson (VP) system. This system consists of two Vlasov equations, one for ions (i) and one for electrons (e), coupled with the Poisson equation for the electrostatic potential:
\begin{eqnarray}
\partial_t f_s + \bsv \cdot \bnb f_s + \frac{q_s}{m_s} \bsE \cdot \bnb_{\bsv} f_s = 0, \quad s = i, e\,, \label{Vlasov_a}\\
\epsilon_0 \Delta \phi = -\sum_{s} q_s \int d^3v\, f_s\,,\label{Poisson_eq_1}\\
\bsE(\bsx,t)=-\bnb\phi(\bsx,t)\,, \label{E-phi}
\end{eqnarray}
where $\Delta$ is the Laplacian operator, $f_s(\bsx,\bsv,t)$ are the distribution functions of the ions ($s=i$) and electrons ($s=e$), $\bsE(\bsx,t)$ is the electric field and $\phi(\bsx,t)$ is the electrostatic potential; $q_i=e$ and $q_e=-e$, with $e$ being the fundamental electric charge; $m_s$ stands for the masses of the ions and electrons; $\epsilon_0$ is the vacuum permittivity, and $\bsx$ and $\bsv$ are the spatial and velocity coordinates, respectively.

The electrostatic potential is a solution to the Poisson equation \eqref{Poisson_eq_1}:
\begin{eqnarray}
    \phi(\bsx,t) = \epsilon_0^{-1}\sum_{s}q_s\int d^3x' d^3v'\, G(\bsx,\bsx') f_s(\bsx',\bsv',t)\,, \label{Green}
\end{eqnarray}
where $G(\bsx,\bsx')$ is the Green function for the Laplacian $\Delta$.

The Vlasov equations \eqref{Vlasov_a} can be formulated as noncanonical Hamilton's equations \cite{Morrison1980b,Morrison1982,Morrison1987,Li2025}, with the following Hamiltonian functional:
\begin{eqnarray}
    \Hc_{_{VP}}=\sum_{s}\int d^3x\,d^3v \, f_s\left( m_s\frac{v^2}{2}+ \frac{q_s}{2}\phi\right) \,, \label{Hamiltonian_VP}
\end{eqnarray}
where $\phi$ is given by \eqref{Green}, and the standard particle Poisson bracket \cite{Morrison1980b,Morrison1982,Marsden1982}:
\begin{eqnarray}
    \{F,G\} = \sum_{s}\int d^3x\, d^3v\, \frac{f_s}{m_s}\left[\frac{\delta F}{\delta f_s},\frac{\delta G}{\delta f_s}\right]_{x,v} \,, \label{Poisson_bracket_VP}
\end{eqnarray}
where $\delta F/\delta f_s$ is the functional derivative of $F$ with respect to $f_s$ and
$$ \big[a,b\big]_{x,v}  = \bnb a\cdot\bnb_{\bsv} b- \bnb b\cdot\bnb_{\bsv} a\,. $$
Vlasov equations \eqref{Vlasov_a} follow from
\begin{eqnarray}
    \partial_t f_s = \{f_s,\Hc_{_{VP}}\}_{_{VP}}\,, \quad s=i,e \,,
\end{eqnarray}
noticing that $$\frac{\delta\Hc_{_{VP}}}{\delta f_s}= \frac{1}{2}m_s v^2+q_s \phi\,, $$
which is the total particle energy.

Finally, it is well known that the bracket \eqref{Poisson_bracket_VP} has the following infinite family of Casimir invariants, i.e. functionals $\mathcal{C}$ satisfying $\{F,\mathcal{C}\}=0\,,$ $\forall F$:
\begin{eqnarray}
    \mathcal{C}_s = \int d^3xd^3v\, \mathscr{C}_s(f_s)\,, \label{C_a}
\end{eqnarray}
where $\mathscr{C}_s$ are arbitrary, well-behaved functions of $f_{s}$.

\subsection{The Vlasov-Amp\`ere system}
Another model used to describe electrostatic evolution of plasmas with negligible  or totally  vanishing magnetic field, is the Vlasov-Amp\`ere system. Now, the Vlasov equations are closed with the Amp\`ere equation for the calculation of the electric field, instead of  the Poisson equation \eqref{Poisson_eq_1}, i.e. we have the following dynamical equation for $\bsE(\bsx,t)$:
\begin{eqnarray}
    \epsilon_0 \partial_t\bsE = -\bsJ\,,
\end{eqnarray}
where
\begin{eqnarray}
    \bsJ(\bsx,t) = \sum_{s} q_s \int d^3v\, f_s \bsv\,, \label{J}
\end{eqnarray}
is the electric current density. The VA system can be formulated as a noncanonical Hamiltonian system with the Hamiltonian:
\begin{eqnarray}
    \Hc_{_{VA}}=\sum_{s}\int d^3x\,d^3v \, f_s m_s \frac{v^2}{2} + \frac{1}{2\epsilon_0}\int d^3x\, |\bsE|^2 \,, \label{Hamiltonian_VA}
\end{eqnarray}
and the bracket:
\begin{eqnarray}
    \{F,G\}_{_{VA}} = \sum_{s}\int d^3x\, d^3v\,\bigg\{ \frac{f_s}{m_s}\left[\frac{\delta F}{\delta f_s},\frac{\delta G}{\delta f_s}\right]_{x,v}\nn\\
    +\frac{q_s}{\epsilon_0 m_s}\left(\frac{\delta G}{\delta f_s}\frac{\delta F}{\delta\bsE} \cdot\bnb_{\bsv} f_s -\frac{\delta F}{\delta f_s} \frac{\delta G}{\delta\bsE}\cdot\bnb_{\bsv}f_s \right) \bigg\}\,. \label{Poisson_bracket_VA}
\end{eqnarray}

The dynamical equations can be cast in the form of the Hamilton equations:
$$ \partial_t \boldsymbol{u} = \{\boldsymbol{u},\Hc_{_{VA}}\}_{_{VA}}\,, \quad \boldsymbol{u} = (f_s,E)\,. $$

 It is important to note that the bracket \eqref{Poisson_bracket_VA} does not, in general, define a Poisson bracket, as it fails to satisfy the Jacobi identity. However, it does satisfy the identity  and thus becomes a true Poisson bracket, if the electric field remains irrotational, a condition required for the validity of the VA system, as mentioned in the introduction. This can be shown by following the proof of the Jacobi identity for the full VM bracket presented in \cite{Morrison2013}, (see Appendix~\ref{app_1}). Note that this condition is trivially satisfied in one-dimensional plasmas, so the VA system and the corresponding Hamiltonian description are always valid in one dimension.  Hence, as noted in the introduction, while we adopt the generic three-dimensional formalism in this subsection to maintain uniform notation with the analysis of the VP system, our treatment of the VA system ultimately concentrates on the 1D case.

The bracket \eqref{Poisson_bracket_VA} has the two infinite families of Casimirs, one is given by \eqref{C_a} and the second one is:
\begin{eqnarray}
    \Cc_{\bsE} &=& \int d^3x\, \mathrm{g}(\bsx) \left( \epsilon_0 \bnb\cdot \bsE -\sum_{s}q_s \int d^3v\, f_{s} \right)\,, \label{C_E_1}
\end{eqnarray}
where $\mathrm{g}(\bsx)$ is an arbitrary function of $\bsx$. Thus, in the VA model, the Poisson equation arises as a consequence of the conservation of $\Cc_{\bsE}$, since $d\Cc_{\bsE}/dt=0$ implies 
$$ \partial_t \left( \epsilon_0 \bnb\cdot \bsE -\sum_{s}q_s \int d^3v\, f_{s} \right)=0\,.$$
Thus, if the Gauss law is initially satisfied, it must be satisfied for all times.

\section{Quasineutrality as a Dirac constraint}
\label{sec_III}

\subsection{The reformulated VP system}

Quasineutrality, i.e. $n_i(\bsx,t) = n_e(\bsx,t)$, arises naturally  when considering large length scales, much larger than the electron Debye length, where electric fields are effectively screened by the particle response. This can be seen by introducing nondimensional quantities:
\begin{eqnarray}
    \tilde{x} = \frac{x}{L}\,, \quad\tilde{v}=\frac{v}{v_{th,e}}\,, \quad \tilde{t} = \frac{t}{L/v_{th,e}}\,, \quad  f_{s} = \frac{f}{n_0/ v_{th,e}^3}\,, \quad \tilde{E} = \frac{E}{k_B T_e/eL}\,,
\end{eqnarray}\
where $L$, $n_0$ are the characteristic length scale and density, respectively, and 
$$ v_{th,e} = \sqrt{\frac{k_B T_e}{m_e}}\,, $$
is the electron thermal velocity. By this normalization the VP system is written in the following non-dimensional form:
\begin{eqnarray}
    \partial_t f_s + \bsv\cdot \bnb f_{s} + \mu_{s} \bsE\cdot\bnb_{\bsv} f_{s}=0\,,\\
    -\lambda^2 \Delta \phi = n_i-n_e\,, \label{poisson_lambda}
\end{eqnarray}
where $\mu_i=\mu = m_e/m_i$, $\mu_e = -1$ and $\lambda=\lambda_{D,e}/L$\,, with $$ \lambda_{D,e} = \sqrt{\frac{\epsilon_0 k_B T_e}{n_0 e^2}}\,,$$ being the electron Debye length. For length scales $L \gg \lambda_{D,e}$, i.e. $\lambda \ll 1$, Poisson equation \eqref{poisson_lambda} implies $n_i=n_e$ where 
\begin{eqnarray}
    n_s = \int d^3v\, f_s\,, \quad s =i,e\,,
\end{eqnarray}
are the particle densities of the two particle species.

A series of papers \cite{Crispel2005a,Crispel2005b, Degond2006,Belaouar2009, Degond2010} considered the asymptotic limit $\lambda \rightarrow 0$, where the Poisson equation cannot close the system anymore as it merely yields $n_i=n_e$. To close the Vlasov equation, those works consider a reformulated VP system  where the Poisson equation is replaced by another elliptic partial differential equation   for the determination of the electrostatic potential, obtained by taking zeroth and first order moments of the Vlasov equation. Manipulating appropriately the moment equations the following equation arises:
\begin{eqnarray}
    -\bnb\cdot \left[ (n_e+\mu n_i+\lambda^2\partial_{tt})\bnb \phi \right] = \bnb\bnb\boldsymbol{\colon} (\bsP_i-\bsP_e)  \,, \label{reformulated_1}
\end{eqnarray}
which allows the calculation of $\phi$ even in the limit $\lambda \rightarrow 0$. In Eq. \eqref{reformulated_1} the tensors $\bsP_s$ are defined as: 
\begin{eqnarray}
    \bsP_s=\int d^3v\, \bsv\bsv f_s\,. \label{P_s}
\end{eqnarray}

\subsection{The Dirac method of constraints}

Here, we follow a different approach, exploiting the Hamiltonian formulation of the model to impose the quasineutrality condition as a Dirac constraint following  the Dirac algorithm for the construction of a generalized Poisson bracket \cite{Dirac1950,Dirac1958,Sundermeyer1982,Morrison2020}. The quasineutrality condition 
\begin{eqnarray}
    \int d^3v \, f_i = \int d^3v \, f_e\,,
\end{eqnarray}
can be seen as a constraint:
\begin{eqnarray}
    \Phi_1(\bsx) = \int d^3v\, \left[f_i(\bsx,\bsv,t)-f_e(\bsx,\bsv,t)\right]=0\,. \label{Phi_1_1}
\end{eqnarray}
A consistency conditions is that the constraint function $\Phi_1(\bsx)$ should be preserved by the dynamics, so  $\{\Phi_1,\Hc\}\approx 0$, where `$\approx$' denotes weak equality, i.e. the equality is satisfied on submanifolds of the phase-space identified by the constraint. To impose this consistency condition, we need to write the constraint function $\Phi_1$ as a phase-space functional, i.e.
\begin{eqnarray}
    \Phi_1(\bsx)= \int d^3x' d^3v\, \delta(\bsx-\bsx')\left[f_i(\bsx',\bsv,t)-f_e(\bsx',\bsv,t)\right]\,, \label{Phi_1_2}
\end{eqnarray}
so that the functional derivatives of $\Phi_1$ are 
\begin{eqnarray}
    \frac{\delta \Phi_1}{\delta f_i} = \delta(\bsx-\bsx')\,, \quad \frac{\delta \Phi_1}{\delta f_e}=-\delta(\bsx-\bsx')\,. \label{deltaF_deltaf}
\end{eqnarray}

\subsubsection{The Vlasov-Poisson system}
In view of \eqref{deltaF_deltaf} and \eqref{Poisson_bracket_VP} we find
\begin{eqnarray}
    \{\Phi_1,\Hc\}_{_{VP}} = -\bnb\cdot \int d^3v\, \bsv(f_i-f_e) = -\bnb\cdot \bsJ\,. 
\end{eqnarray}
Hence, the requirement $\{\Phi_1,\Hc\}_{_{VP}}\approx 0$ results in a secondary constraint:
\begin{eqnarray}
    \Phi_2(\bsx) = \int d^3x'\, \delta(\bsx-\bsx') \bnb\cdot \bsJ = \int d^3x'\, \delta(\bsx-\bsx') \bnb\cdot\int d^3v\,\bsv(f_i-f_e)=0\,, \label{Phi_2}
\end{eqnarray}
 a condition that also arises in the analysis of \cite{Burby2025}. The consistency condition $\{\Phi_2,\Hc\}_{_{VP}}\approx0$ can verified that holds true in view of the two unconstrained Vlasov equations \eqref{Vlasov_a}. 

We can verify that $\Phi_1$ and $\Phi_2$ are second-class constraints, i.e. the quantity $\{\Phi_1,\Phi_2\}_{VP}$ is non-vanishing, so, we can form a non-singular, antisymmetric constraint matrix of the form:
\begin{eqnarray}
    C=\begin{pmatrix}
        C_{11} && C_{12}\\
        C_{21} && C_{22}
    \end{pmatrix}\,, \label{constr_matrix}
\end{eqnarray}
where $C_{jk}(\bsx,\bsx') = \{\Phi_j(\bsx),\Phi_k(\bsx')\}_{VP}$ with $j,k=1,2$. The details of the calculation of the elements $C_{jk}$ of the constraint matrix are presented in Appendix  \ref{app_2}. These elements are:
\begin{eqnarray}
    C_{11}=0\,, \quad C_{12}(\bsx,\bsx')=-C_{21}(\bsx,\bsx') = \Lc \delta(\bsx'-\bsx)\,, \nn\\
    C_{22}(\bsx,\bsx')= \bnb\cdot\left[(\bsM\cdot\bnb)\bnb \delta(\bsx'-\bsx)+\bsM \Delta \delta(\bsx'-\bsx)+\bnb\delta(\bsx'-\bsx)\cdot\bnb\bsM\right]\,, \label{C_elements}
\end{eqnarray}
where
\begin{eqnarray}
    \Lc:= \bnb\cdot\left[ \left(\frac{n_i}{m_i}+\frac{n_e}{m_e}\right)\bnb \right]\,,
\end{eqnarray}
is a self-adjoint elliptic operator and 
\begin{eqnarray}
    \bsM(\bsx,t) = \int d^3v\, \bsv\left(\frac{f_i}{m_i}+\frac{f_e}{m_e}\right)\,. \label{M}
\end{eqnarray}
In order to eliminate the second-class constraints, Dirac defined a new bracket
algebra on the phase space such that the bracket of any phase space function with a
constraint vanishes, thus the constraints become Casimir invariants of the new bracket. The Dirac bracket of two functionals $F$ and $G$ on the phase space  is defined as follows:
\begin{eqnarray}
    \{F,G\}^* = \{F,G\}- \int\int d^3x d^3x' \{F,\Phi_j(\bsx)\}C^{-1}_{jk}(\bsx,\bsx')\{\Phi_{k}(\bsx'),G\}\,, \label{Dirac_bracket_1}
\end{eqnarray}
where $\{F,G\}$ is the standard Poisson bracket and $C^{-1}_{jk}$ are the inverse matrix elements calculated by
\begin{eqnarray}
    \int d^3x'' C_{j\ell}(\bsx,\bsx'')C^{-1}_{\ell k}(\bsx'',\bsx') = \delta_{jk} \delta(\bsx-\bsx')\,.
\end{eqnarray}
After some straightforward manipulations we find:
\begin{eqnarray}
    C_{22}^{-1}=0\,, \quad C_{12}^{-1}(\bsx,\bsx') = -C_{21}^{-1}(\bsx,\bsx')=-\Lc^{-1}\delta(\bsx-\bsx')\,, \nn \\
    C_{11}^{-1}(\bsx,\bsx') = \Lc^{-1}\Lc'^{-1}\bnb\cdot\left[(\bsM\cdot\bnb)\bnb\delta(\bsx'-\bsx)+\bsM\Delta\delta(\bsx'-\bsx)+\bnb\delta(\bsx'-\bsx)\cdot\bnb\bsM\right]\,,\label{inverse_C_1}
\end{eqnarray}
where $\Lc^{-1}$ is the inverse of $\Lc$. To derive these elements, we have used that if $\Lc$ is a self-adjoint and invertible operator on $L^2$ Hilbert space (so that its inverse $\Lc^{-1}$ exists on $L^2$), then its inverse  $\Lc^{-1}$ is also self-adjoint. 

Substituting the elements $\Cc^{-1}_{jk}$ from \eqref{inverse_C_1} into the general form of the field-theoretic Dirac bracket \eqref{Dirac_bracket_1}, and after some tedious manipulations, we arrive at the following bracket:
\begin{eqnarray}
    \{F,G\}^\star = \{F,G\}_{_{VP}}-\int d^3x\, \Big\{ \bnb\left(\Lc^{-1}  \Upsilon_F\right) \bsM \boldsymbol{:} \bnb\bnb\left(\Lc^{-1} \Upsilon_G\right) \nn \\
    - \bnb\left(\Lc^{-1}  \Upsilon_G\right) \bsM \boldsymbol{:} \bnb\bnb\left(\Lc^{-1} \Upsilon_F\right)+\Omega_G\Lc^{-1} \Upsilon_F-\Upsilon_G\Lc^{-1}\Omega_F \Big\}\,. \label{Dirac_bracket_2}
\end{eqnarray}
where
\begin{eqnarray}
    \Upsilon_X &:=& \bnb\cdot\int d^3v\, \left[  \frac{\delta X}{\delta f_i}\bnb_{\bsv}\left( \frac{f_i}{m_i}\right) - \frac{\delta X}{\delta f_e}\bnb_{\bsv}\left( \frac{f_e}{m_e}\right)\right]\,, \label{Upsilon_X} \\
    \Omega_X &:=& \bnb\cdot \int d^3v\, \bsv\left(\left[\frac{\delta X}{\delta f_i},\frac{f_i}{m_i}\right]_{x,v}-\left[\frac{\delta X}{\delta f_e},\frac{f_e}{m_e}\right]_{x,v}\right)\,, \quad X=F,G\,. \label{Omega_X_VP}
\end{eqnarray}
and $\{F,G\}_{_{VP}}$ is the standard Poisson bracket \eqref{Poisson_bracket_VP}. One can show that $\Cc_s$, $\Phi_1$ and $\Phi_2$ given by \eqref{C_a}, \eqref{Phi_1_2} and \eqref{Phi_2}, respectively, are Casimir invariants of the bracket \eqref{Dirac_bracket_2}. Hence, the second-class constraints have been incorporated into the dynamics of the system, and their conservation is guaranteed by the structure of the Dirac bracket \eqref{Dirac_bracket_2}, which replaces the original Poisson bracket \eqref{Poisson_bracket_VP}.

\subsubsection{The Vlasov-Amp\`ere system}
For the VA system we require that the constraint $\Phi_1$ satisfies the consistency condition $\{\Phi_1,\Hc\}_{_{VA}}\approx 0$ where the Poisson bracket is now given by \eqref{Poisson_bracket_VA}. We can easily show that, by this condition, the same secondary constraint \eqref{Phi_2} arises, and we can employ the same steps as in the VP case  to derive the corresponding Dirac bracket in the VA case. After repeating the procedure we find that the Dirac bracket for the VA system is formally identical to \eqref{Dirac_bracket_2} with the difference that $\{F,G\}_{VP}$ is replaced by $\{F,G\}_{VA}$ given by \eqref{Poisson_bracket_VA} and the quantities $\Omega_{X}$ are now given by:
\begin{eqnarray}
    \Omega_X = \bnb\cdot \int d^3v\, \left[\bsv\left(\left[\frac{\delta X}{\delta f_i},\frac{f_i}{m_i}\right]_{x,v}-\left[\frac{\delta X}{\delta f_e},\frac{f_e}{m_e}\right]_{x,v}\right)+\frac{e}{\epsilon_0}\frac{\delta X}{\delta\bsE} \left(\frac{f_i}{m_i}+\frac{f_e}{m_e}\right)\right]\,, \nn\\
     X=F,G\,. \label{Omega_X_VA}
\end{eqnarray}
Casimir invariants of the new bracket are $\Cc_s$, the two second-class constraints $\Phi_1$ and $\Phi_2$ given by \eqref{C_a}, \eqref{Phi_1_2} and \eqref{Phi_2}, respectively, and 
\begin{eqnarray}
\Cc_{\bsE} = \int d^3x\, \mathrm{g}(\bsx)\bnb\cdot\bsE\,, \label{C_E_2} 
\end{eqnarray}
which is consistent with \eqref{C_E_1}, since the quasineutrality condition has been imposed.
\subsection{Constrained dynamics}
For the VP system, the constrained dynamical equations that respect the preservation of the constraints $\Phi_1$ and $\Phi_2$ given by \eqref{Phi_1_2} and \eqref{Phi_2}, respectively, are recovered from the Hamilton-Dirac equations:
\begin{eqnarray}
    \partial_t f_s=\{f_s,\Hc_{_{VP}}\}^\star\,, \label{Hamilton-Dirac_VP_1}
\end{eqnarray}
where $\Hc_{_{VP}}$ is given by \eqref{Hamiltonian_VP} and $\{F,G\}^\star$ is given by \eqref{Dirac_bracket_2} with $\Upsilon_X$ and $\Omega_X$ specified in Eqs.~\eqref{Upsilon_X} and \eqref{Omega_X_VP}, respectively.  For the VA system, the corresponding constrained equations are: 
\begin{eqnarray}
    \partial_t f_s = \{f_s,\Hc_{_{VA}}\}^\star\,, \quad s =i,e \label{Hamilton-Dirac_VA_f_1}\\
    \partial_t \bsE = \{\bsE,\Hc_{_{VA}}\}^\star\,, \label{Hamilton-Dirac_VA_E_1}
\end{eqnarray}
 with $\Hc_{_{VA}}$ given by \eqref{Hamiltonian_VA} and $\{F,G\}^\star$ by \eqref{Dirac_bracket_2} where $\Upsilon_X$ and $\Omega_X$ are given by \eqref{Upsilon_X} and \eqref{Omega_X_VA}, respectively. Both equations \eqref{Hamilton-Dirac_VP_1} and \eqref{Hamilton-Dirac_VA_f_1} yield the following system of constrained Vlasov equations:
\begin{eqnarray}
    \partial_t f_s &=& - \bsv\cdot\bnb f_s - \frac{q_s}{m_s}\bsE\cdot\bnb_{\bsv} f_s \pm \frac{1}{m_s}\bigg\{ \bnb\alpha \cdot \bnb f_s -\left[(\bsv\cdot\bnb) \bnb \alpha - \bnb \beta + \bnb \gamma\right]\cdot \bnb_{\bsv} f_{s} \nn\\
    &&+q_s\bnb_{\bsv} f_s\cdot\bnb\Lc^{-1}\bnb\cdot\int d^3v\,\bsE\left(\frac{f_i}{m_i}+\frac{f_e}{m_e}\right) \bigg\}\,, \label{Hamilton-Dirac_f_1}
\end{eqnarray}
where $+$ corresponds to the ion equation and $-$ to the electron equation. In Eq. \eqref{Hamilton-Dirac_f_1} the quantities $\alpha$, $\beta$, $\gamma$ are:
\begin{eqnarray}
    \alpha= \Lc^{-1} \bnb\cdot\int d^3v\, \bsv(f_i-f_e) = e^{-1}\Lc^{-1} \bnb\cdot \bsJ\,, \\
    \beta = \Lc^{-1}\bnb\cdot\left[(\bsM\cdot\bnb)\bnb\alpha+\bsM\Delta\alpha+(\bnb\alpha\cdot\bnb)\bsM\right]\,, \\
    \gamma=\Lc^{-1}\bnb\cdot\int d^3v\, \bsv\bsv\cdot\bnb(f_i-f_e)=\Lc^{-1}\bnb\bnb\boldsymbol{:}(\bsP_i-\bsP_e) \,;
\end{eqnarray}
hence, they can be computed upon solving the following elliptic equations:
\begin{eqnarray}
    \Lc \alpha &=& e^{-1}\bnb\cdot\bsJ \,, \label{Lalpha}\\
    \Lc\beta &=&\bnb\cdot\left[(\bsM\cdot\bnb)\bnb\alpha+\bsM\Delta\alpha+(\bnb\alpha\cdot\bnb)\bsM\right]\,, \label{Lbeta}\\
    \Lc \gamma&=&\bnb\bnb\boldsymbol{:}(\bsP_i-\bsP_e)\,. \label{Lgamma}
\end{eqnarray}
The electric field is given by $\bsE=-\bnb\phi$ for the general VP system, while for the VA system, Eq.~\eqref{Hamilton-Dirac_VA_E_1} yields:
\begin{eqnarray}
    \epsilon_0 \partial_t \bsE =- \bsJ +e\left(\frac{n_i}{m_i}+\frac{n_e}{m_e}\right)\bnb\alpha\,. \label{constr_ampere}
\end{eqnarray}
Taking the divergence of this equation, yields $\partial_t (\nabla \cdot \bsE) = 0$, in agreement with the Casimir invariant \eqref{C_E_2}, since \eqref{Lalpha} implies that the two terms on the right-hand side of \eqref{constr_ampere} form a divergence-free vector. Obviously, in one dimension, the condition $\nabla \times \bsE = 0$ holds automatically, whereas in higher dimensions \eqref{constr_ampere} implies that this condition is satisfied provided the current density $\bsJ$ obeys $
	\bnb \times \bsJ = e \, \bnb \left(\frac{n_i}{m_i} + \frac{n_e}{m_e}\right) \times \bnb \alpha$.
However, as previously discussed,  in dimensions greater than one the VA system must be replaced by the full Vlasov--Maxwell system.

Now, let us consider the VP case, and further notice that upon substituting $\bsE = -\bnb \phi$ into \eqref{Hamilton-Dirac_f_1}, the last term becomes:
$$ \pm \frac{q_s}{m_s}  (\bnb_{\bsv} f_s)\cdot\bnb\Lc^{-1}\bnb\cdot\int d^3v\,\bsE\left(\frac{f_i}{m_i}+\frac{f_e}{m_e}\right)=\mp \frac{e} {m_s}\bnb_{\bsv}f_s\cdot \bnb \Lc^{-1}\Lc \phi= \mp \frac{q_s}{m_s} \bnb \phi \cdot \bnb_{\bsv}f_s\,.$$
Substituting this last equation in \eqref{Hamilton-Dirac_f_1} we see that the electric field term is effectively eliminated from the constrained Vlasov equations, and therefore these two equations comprise a closed system, since the new advection fields
\begin{eqnarray}
\bsxi :=\bnb\alpha\,, \quad\bszeta:= \bnb \beta\,, \quad \bseta:=\bnb\gamma\,,
\end{eqnarray}
depend merely on $f_i$ and $f_e$. 

Note also that, along the same lines, we can find an analogous result for the one-dimensional VA case, since the electric field, being a continuous function, can be written as the gradient of a differentiable function. Therefore, the final system of QN Vlasov equations, both in the VP and the one-dimensional VA case, takes the form:
\begin{eqnarray}
       \partial_t f_s  + \bsv\cdot\bnb f_s - \frac{q_s}{em_s}\left\{ \bsxi \cdot \bnb f_s -\left[(\bsv\cdot\bnb) \bsxi - \bszeta + \bseta \right]\cdot \bnb_{\bsv} f_{s} \right\}=0\,, \quad s=i,e\,. \label{Hamilton-Dirac_f_2}
\end{eqnarray}
For the advection of the distribution functions $f_i$ and $f_e$, we are interested in calculating the fields $\bsxi$, $\bszeta$ and $\bseta$, which according to \eqref{Lalpha}--\eqref{Lgamma} are
\begin{eqnarray}
    \bnb \cdot\left[(n_e+\mu n_i) \bsxi\right]&=&\frac{m_e}{e}\bnb\cdot\bsJ\,,\\
    \bnb\cdot\left[(n_e+\mu n_i)\bszeta\right] &=& m_e \bnb\cdot \left[(\bsM\cdot\bnb)\bsxi+\bsM\bnb\cdot\bsxi+(\bsxi\cdot\bnb)\bsM\right]\,,\\
    \bnb\cdot\left[(n_e+\mu n_i) \bseta\right] &=& m_e \bnb\bnb\boldsymbol{:}(\bsP_i-\bsP_e)\,. \label{}
\end{eqnarray}
Note that the indices $i$ and $e$ in the particle densities are retained, even though we have imposed $n_i=n_e$ as a Dirac constraint. This is because the Dirac algorithm incorporates the constraint into the dynamics by modifying the bracket structure, such that the constraint appears as a Casimir invariant of the form \eqref{Phi_1_2}. As a result, the constrained dynamics does not explicitly enforce $n_i=n_e$, but rather enforces the invariance of $\Phi_1$. This implies that the difference $n_i-n_e$ is conserved over time. Therefore, only if the system is initialized with $n_i-n_e=0$ (i.e., it satisfies quasineutrality), we can write $n_i=n_e=n$.

In this framework, the Poisson equation of the VP system, and the Amp\`ere equation of the 1D VA system become irrelevant and superfluous, since the electric field does not participate in the advection of the distribution functions. The new advection fields are
$$
    \bsv  - \frac{q_s}{em_s}\bsxi\,,
$$
in the physical (configuration) space and 
$$
\frac{q_s}{em_s}[(\bsv\cdot\bnb)\bsxi-\bszeta+\bseta]\,,
$$
in the velocity space.

\section{Numerical results}
\label{sec_IV}
In this section, we provide a numerical verification of the Dirac-constrained formulation developed in Section \ref{sec_III}. Through complementary tests, we demonstrate two primary outcomes. First, we verify the correct algorithmic implementation by showing that the Dirac constraints, namely, the fixed charge-density distribution ($\Phi_1$) and the current incompressibility ($\Phi_2$), are preserved by the constrained system. Second, the simulations illustrate how QN (QN) dynamics deviate from standard Vlasov-Poisson (VP) dynamics even when starting from identical initial conditions, thereby highlighting how the enforced constraints reshape kinetic evolution. Together, these tests validate the theoretical framework and offer insight into the dynamical consequences of imposing quasineutrality at the kinetic level.

\subsection{The non-dimensional  1D-1V system}

Let us now consider the QN Vlasov system in the one-dimensional spatial and velocity coordinate setting (1D-1V), where the distribution function depends on a single spatial coordinate $x$ and a single velocity coordinate $v$. We also introduce the following normalized quantities:
\begin{eqnarray}
    \tilde{x} = \frac{x}{\lambda_{D,e}}\,, \quad \tilde{v} = \frac{v}{v_{th,e}}\,,  \quad \tilde{t} = \omega_{p,e} t\,, \quad \tilde{f} = \frac{f}{n_0/v_{th,e}}\, , \nn\\
    \quad \tilde{\xi} = \frac{\xi}{m_e v_{th,e}}\,, \quad \tilde{\zeta} = \frac{\zeta}{m_e v_{th,e}^2/\lambda_{D,e}}\,, \quad \tilde{\eta} = \frac{\eta}{m_e v_{th,e}^2/\lambda_{D,e}}
\end{eqnarray}
with 
$$\omega_{p,e}=\sqrt{\frac{n_0e^2}{\epsilon_0m_e}}\,,$$
being the electron plasma frequency, in order to write the 1D-1V counterparts of Eqs. \eqref{Hamilton-Dirac_f_2} in non-dimensional form:
\begin{eqnarray}
    \partial_tf_s+v \partial_x f_s -\mu_s \left[ \xi \partial_x f_s-(v\partial_x\xi-\zeta+\eta) \partial_vf_s\right]=0\,, \quad s=i,e\,, \label{Vlasov_1D1V_f_1}
\end{eqnarray}
and $\mu_i = \mu =m_e/m_i$, $\mu_e =-1$. The functions $\xi$, $\zeta$ and $\eta$ satisfy:
\begin{eqnarray}
    &&\partial_x\left[(n_e+\mu n_i)\xi\right] = \partial_x J\,, \label{dxi_dx}\\
   && \partial_x\left[(n_e+\mu n_i)\zeta\right] = \partial_x\left(2 M\partial_x\xi +\xi \partial_x M\right)\,, \label{dzeta_dx}\\
   && \partial_x\left[(n_e+\mu n_i)\eta\right] = \partial_{xx} (P_i-P_e)\,, \label{deta_dx}
\end{eqnarray}
where $J$, $M$ and $P_s$ are the scalar one-dimensional counterparts of $\bsJ$, $\bsM$ and $\bsP_s$ given by \eqref{J}, \eqref{M} and \eqref{P_s}, respectively. Integrating equations \eqref{dxi_dx}--\eqref{deta_dx} with respect to $x$ and assuming that $n_e+\mu n_i >0$ $\forall x$, we find:
\begin{eqnarray}
    \xi = \frac{J+c_1}{n_e+\mu n_i} \,, \quad \zeta =\frac{2 M\partial_x\xi +\xi \partial_x M +c_2}{n_e+\mu n_i} \,, \quad \eta = \frac{\partial_x (P_i-P_e) +c_3}{n_e+\mu n_i}\,, \label{xi_zeta_eta_1}
\end{eqnarray}
where $c_1,c_2,c_3$ are integration constants. These constants must be independent of time in order to preserve the conservation properties of the Vlasov system. Their values may be fixed by enforcing suitable boundary conditions; however, for periodic boundary conditions these constants remain undetermined, so without loss of generality, we set $c_1=c_2=c_3=0$, to avoid introducing potential non-physical shifts in the distribution function.

\subsection{Solution with the semi-Lagrangian method}
We solve Eqs. \eqref{Vlasov_1D1V_f_1} using a semi-Lagrangian method on a structured Eulerian grid $(x_j, v_k) = (j \Delta x, k \Delta v)$, with $j = 0, 1, \dotsc, N_x$ and $k = 0, 1, \dotsc, N_v$. The time integration is performed using a Strang splitting \cite{Strang1968}, similar to the classical Cheng--Knorr scheme \cite{Cheng1976, Sonnendrucker1999}.

To apply the splitting, we decompose the Vlasov equation for each species into a sequence of simpler subproblems. Each corresponds to a term of the operator $\Dc_s$ in $\partial_t f_s +\Dc_s f_s=0$, that governs the transport of the distribution function $f_s$. This operator is given by:
$$ \Dc_s = \Dc_s^x + \Dc_s^{v}\,,$$
with
$$\Dc_s^x:=(v - \mu_s \xi)\partial_x\,, \quad \Dc_s^{v}:=\mu_s (\eta - \zeta+v\partial_x \xi ) \partial_v \,.$$
This decomposition yields the following subsystems describing different physical effects:
\begin{align}
&\text{Advection in } x: \quad
\partial_t f_s + \Dc_s^x f_s= 0\,, \label{splitted_1} \\
&\text{Advection and shearing in } v: \quad
\partial_t f_s + \Dc_s^{v} f_s= 0\,. \label{splitted_2}
\end{align}
In the  semi-Lagrangian method each subsystem is solved along its corresponding characteristic equations:
\begin{eqnarray}
\frac{dX}{dt} &=&  (V-\mu_s\xi(X))\,, \label{char_1} \\
    \frac{dV}{dt} &=& \mu_s(\eta(X)-\zeta(X)+V\partial_x \xi(X))\,. \label{char_2}
\end{eqnarray}
Discretizing time as $t^n = n \Delta t$, a full time step $\Delta t$ in the advection of $f_s$ according to \eqref{splitted_1} and \eqref{splitted_2}, is approximated by the following Strang splitting:
\begin{eqnarray}
e^{\Delta t (\Dc_s^x + \Dc_s^{v})} 
\approx 
e^{\frac{\Delta t}{2} \Dc_s^{x}}
e^{\Delta t \Dc_s^v}
e^{\frac{\Delta t}{2} \Dc_s^{x}}\,. \label{Strang_splitting}
\end{eqnarray}
Thus, both Eq.~\eqref{char_1} and Eq.~\eqref{char_2} are advanced independently, each as part of the composition of flows defined in \eqref{Strang_splitting}.

In the semi-Lagrangian method we essentially determine the starting point of the characteristic curve $(X^n, V^n)$ ending at the grid point $(X^{n+1}, V^{n+1})$, and set
$f(X^{n+1}, V^{n+1}, t^{n+1}) = f(X^n, V^n, t^n)\,,$
since $f$ is conserved along characteristics. As $(X^n, V^n)$ generally does not lie on the grid, we use cubic interpolation from nearby points to evaluate $f(X^n, V^n)$. Hence, in order to find the starting point $(X^n, V^n)$ we need to solve the characteristic equations \eqref{char_1}--\eqref{char_2} backwards in time. Notice that the characteristic equation \eqref{char_1} is a nonlinear ODE for $X(t)$ and generally a nonlinear solver should be invoked, or a very small time-step should be used. On the other hand, employing the method of integrating factor we can find an exact solution to \eqref{char_2}:
$$V(t + \Delta t) = V(t)\, e^{\mu_s\, \partial_x \xi\, \Delta t}+ \frac{\eta-\zeta}{\partial_x \xi}\left(e^{\mu_s\partial_x \xi \Delta t}-1\right)\,,$$
so in the velocity step, we can compute the starting velocity as 
$$V^n = V^{n+1}e^{-\mu_s \partial_x \xi\,\Delta t} - \frac{\eta-\zeta}{\partial_x \xi} \left(1-e^{-\mu_s \partial_x \xi \,\Delta t}\right) \,.$$
One can see that 
\begin{eqnarray}
    \lim_{\partial_x\xi\rightarrow 0} V^n = V^{n+1}- \mu_s(\eta-\zeta) \Delta t\,,
\end{eqnarray}
hence, if $\partial_x\xi=0$, the starting velocity is simply $V^n=V^{n+1}-\mu_s (\eta-\zeta) \Delta t$.

Thus, in practice, the Strang splitting \eqref{Strang_splitting} is employed by performing the following steps:
\begin{enumerate}

\item Half-step advection in $x$: $$f_s'= f_s\left(x-(v-\mu_s\xi)\Delta t/2,v\right)\,,$$
where $\xi$ is computed by $f_s$.
\item Full-step advection and shearing in $v$:  $$f_s''(x,v) =  f_s'\left(x,ve^{-\mu_s \partial_x \xi'\,\Delta t} - \frac{\eta'-\zeta'}{\partial_x \xi'} \left(1-e^{-\mu_s \partial_x \xi' \,\Delta t}\right) \right)$$
where $\xi'$, $\zeta'$ and $\eta'$  are computed from $f_s'$.

\item Half-step advection in $x$: $$f_s'''= f_s''\left(x-(v-\mu_s\xi'')\Delta t/2,v\right)\,,$$
where $\xi''$ is computed by $f_s''$.

\end{enumerate}

To smooth out spurious oscillations in the particle density profile which may develop in regions where fine-scale structures emerge, we found it effective to apply a Savitzky-Golay filter \cite{Schafer2011} of polynomial order 2 to the distribution functions. This filter minimizes the least-squares error, fitting a second-order polynomial to successive frames of high-frequency data. Thus, this procedure suppresses high-frequency oscillations without overly damping physical features.

The interpolation and filtering steps break energy conservation and the symplectic structure of the continuous dynamics, leading to the small but nonzero drifts of invariants observed in the next subsection. In addition, the standard Cheng–Knorr type semi-Lagrangian splitting is employed here in an operational rather than a geometrical manner. Thus, although it is Hamiltonian for the ordinary Vlasov–Poisson system, it is not designed to preserve the Hamiltonian or the underlying Dirac bracket structure of the constrained QN system. To improve both conservation and long-time stability, an important direction for future work is the development of numerical schemes that respect the modified Hamiltonian structure induced by the Dirac bracket. This includes constructing energy-conserving or, more broadly, structure-preserving strategies within the semi-Lagrangian framework (e.g., \cite{Crouseilles2010, Crouseilles2015, Liu2023}) that aim to retain the geometric properties of the continuous flow at the discrete level. The design of such structure-preserving algorithms for the constrained QN Vlasov model lies beyond the scope of the present study but will be pursued in subsequent work.

\subsection{Simulation setup}

\subsubsection{Electron-proton plasma ($\mu\ll1$)}
We initialize the simulation with two counter-streaming plasma beams having equal ion and electron densities, perturbed by a cosine modulation around the characteristic density $n_0=1$ and different drift velocities $V_e$ and $V_i$:
\begin{eqnarray}
    f_e = \frac{1}{2\sqrt{2\pi}} \left(e^{-(v-V_e)^2/2}+e^{-(v+V_e)^2/2}\right)\left[1+\epsilon\, cos\left(\frac{2k\pi x}{L}+\pi\right)\right] + \delta\frac{ve^{-v^2/2}}{\sqrt{2\pi}}\,,\label{f_e}\\
    f_i = \frac{1}{2\tau\sqrt{2\pi}} \left(e^{-(v-V_i)^2/2\tau^2}+e^{-(v+V_i)^2/2\tau^2}\right)\left[1+\epsilon\, cos\left(\frac{2k\pi x}{L}+\pi\right)\right]\,. \label{f_i}
\end{eqnarray}
Here, $\tau= v_{th,i}/v_{th,e}$ is the ratio of the ion to the electron thermal velocities; $\epsilon$ is a small perturbation parameter; $\delta$ controls the constant current density; $L$ is the size of the periodic box measured in electron Debye lengths $\lambda_{D,e}$ and $k$ is an integer indicating the number of density peaks within the box $L$. Here we select $L=150$, $\tau=1$, $\epsilon=0.05$, $\delta=0$ and $k=3$. The drift velocities are $V_e=2$ and $V_i=0.5$. To simulate the evolution of the distribution functions $f_i$ and $f_e$ we consider a $120\times 120$ grid discretizing the simulation box, so that $\Delta x \sim \lambda_{D,e}$ while in velocity space the computational domain spans from $-4\pi$ to $4\pi$, measured in electron thermal velocity units. The electron to ion mass ratio considered here is $\mu=1/1836$, which corresponds to an eletron-proton plasma. We also considered $\mu=1$ for electron-positron plasma. 

\begin{figure}[h!]
    \centering
    \includegraphics[width=0.5\linewidth]{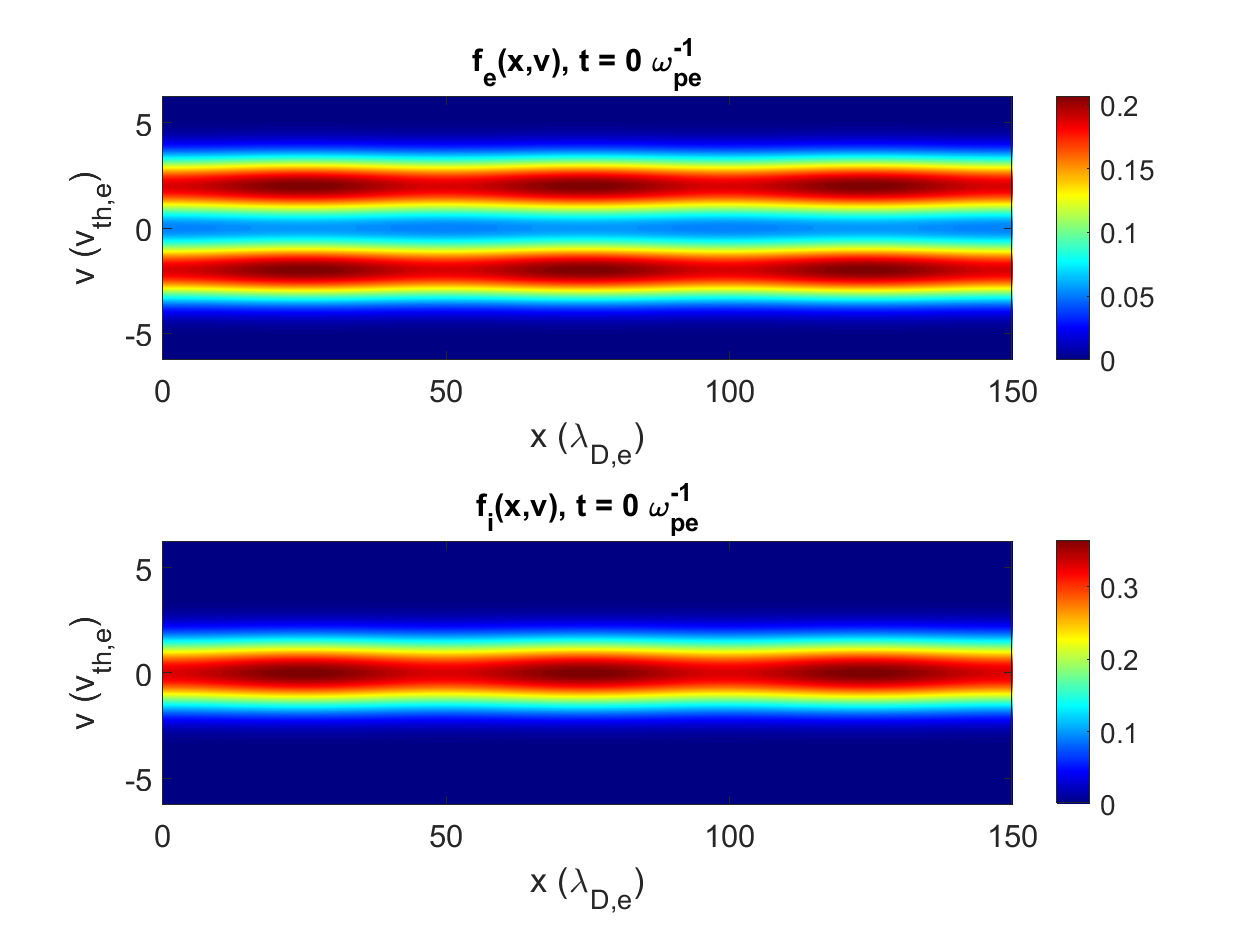}
    \caption[font=small,labelfont=bf]{Initial conditions of the distribution functions $f_e$ and $f_i$ in phase space. The electron distribution forms two separated beams since $V_e$ is large enough in \eqref{f_e}, whereas the two ion beams are so close to each other effectively forming a single beam with zero macroscopic velocity. This broad, centralized ion beam was found to favor numerical stability over longer simulation times.}
    \label{fig_1}
\end{figure}

\begin{figure}[h!]
    \centering
    \includegraphics[width=0.45\linewidth]{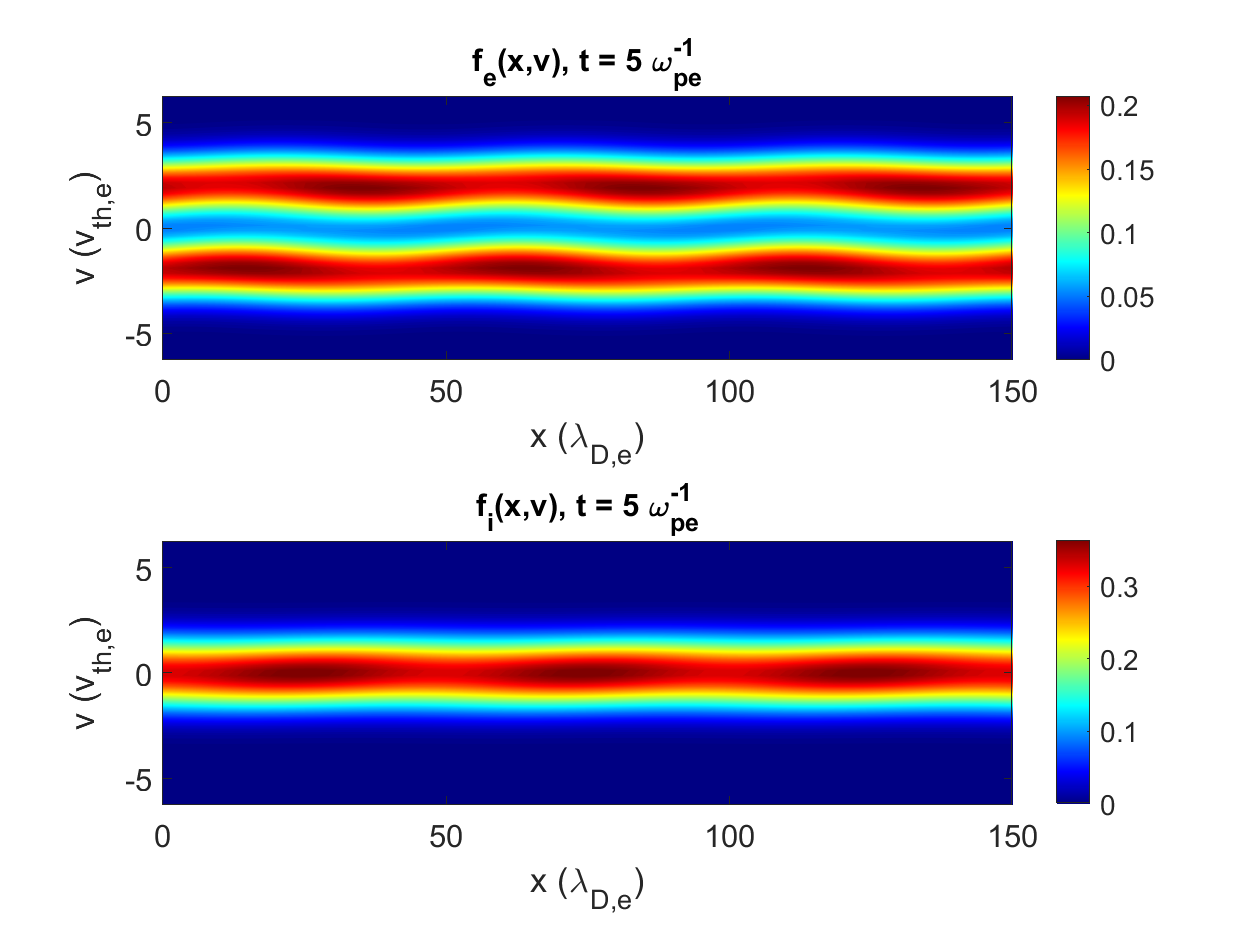}     \includegraphics[width=0.45\linewidth]{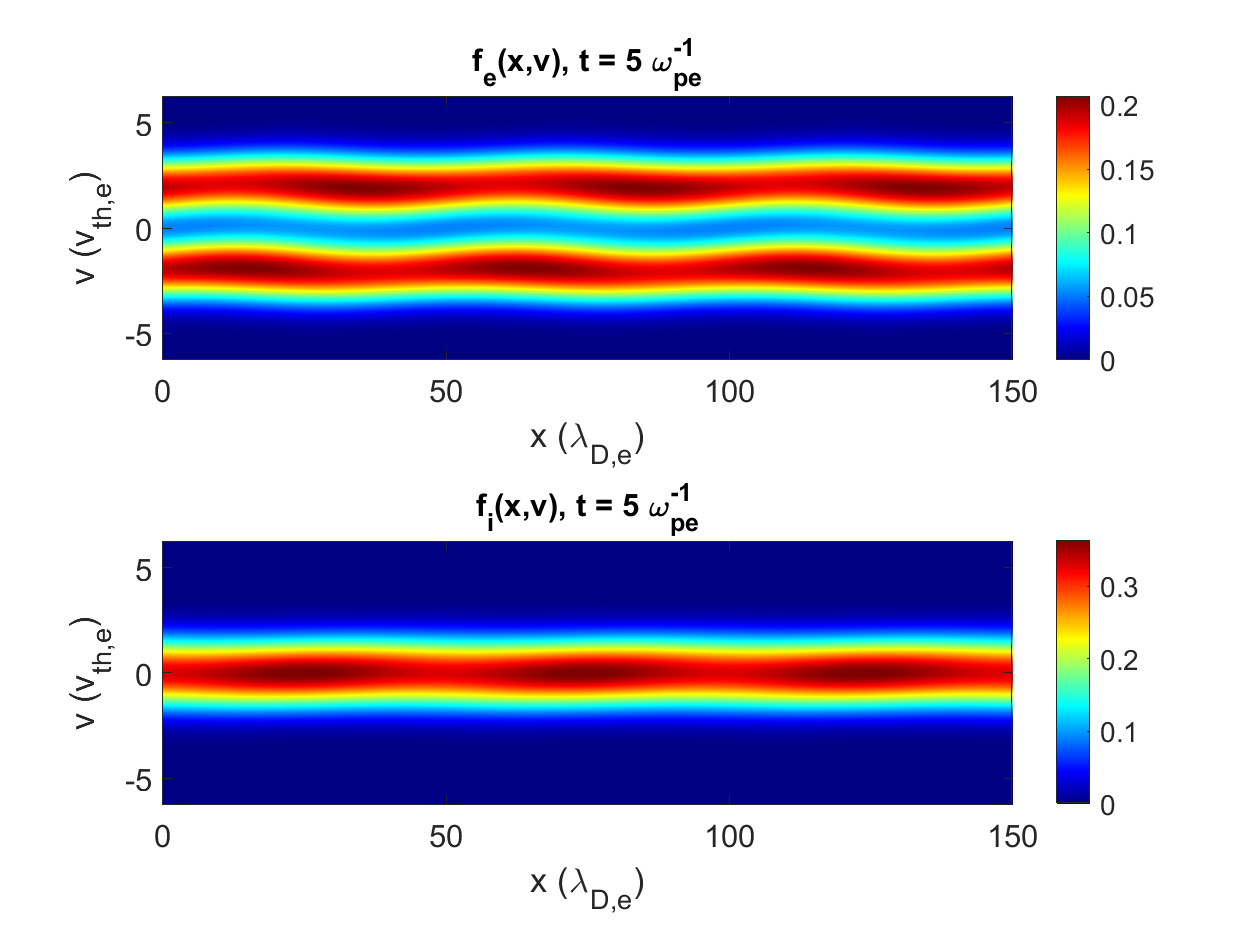}
        \includegraphics[width=0.45\linewidth]{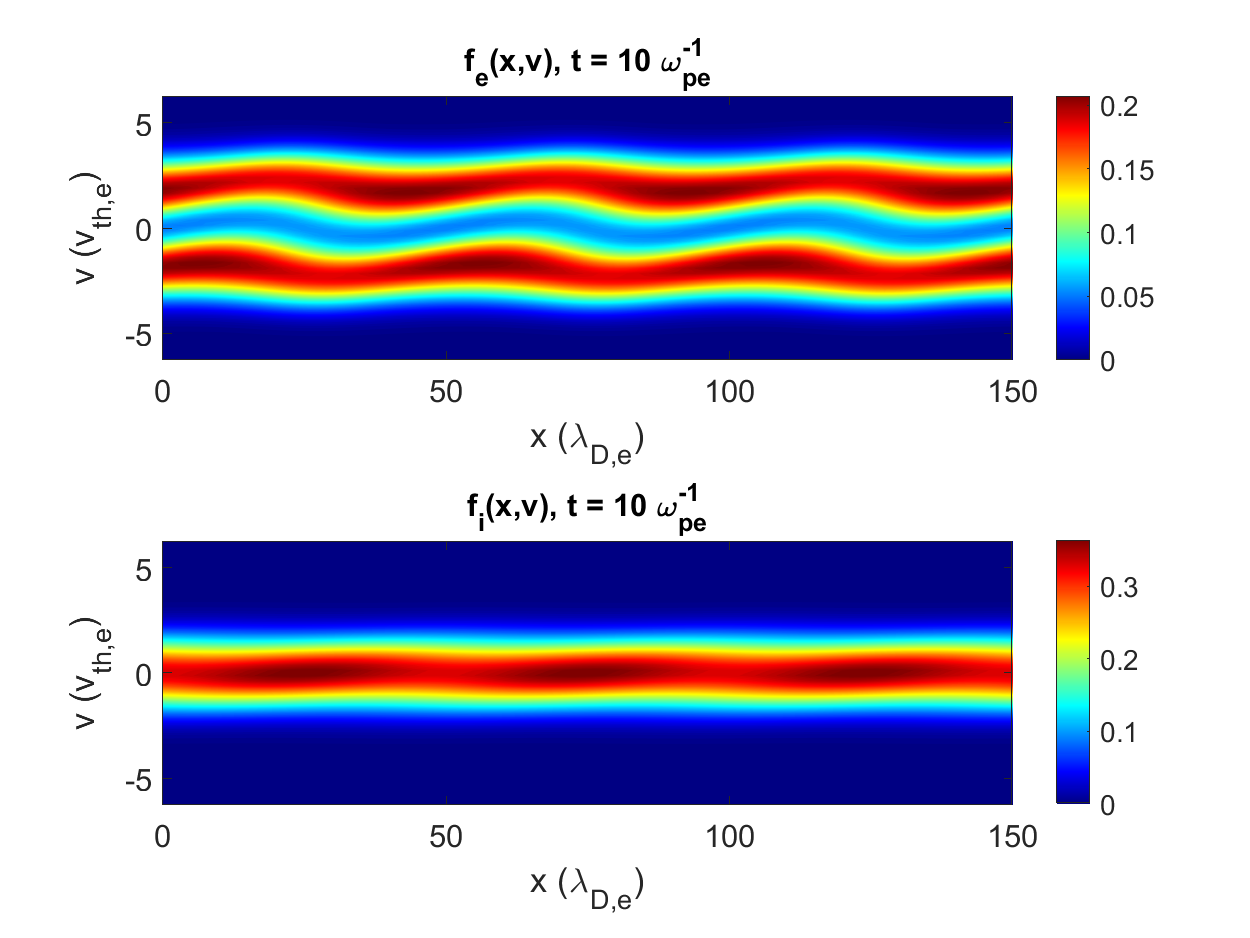}     \includegraphics[width=0.45\linewidth]{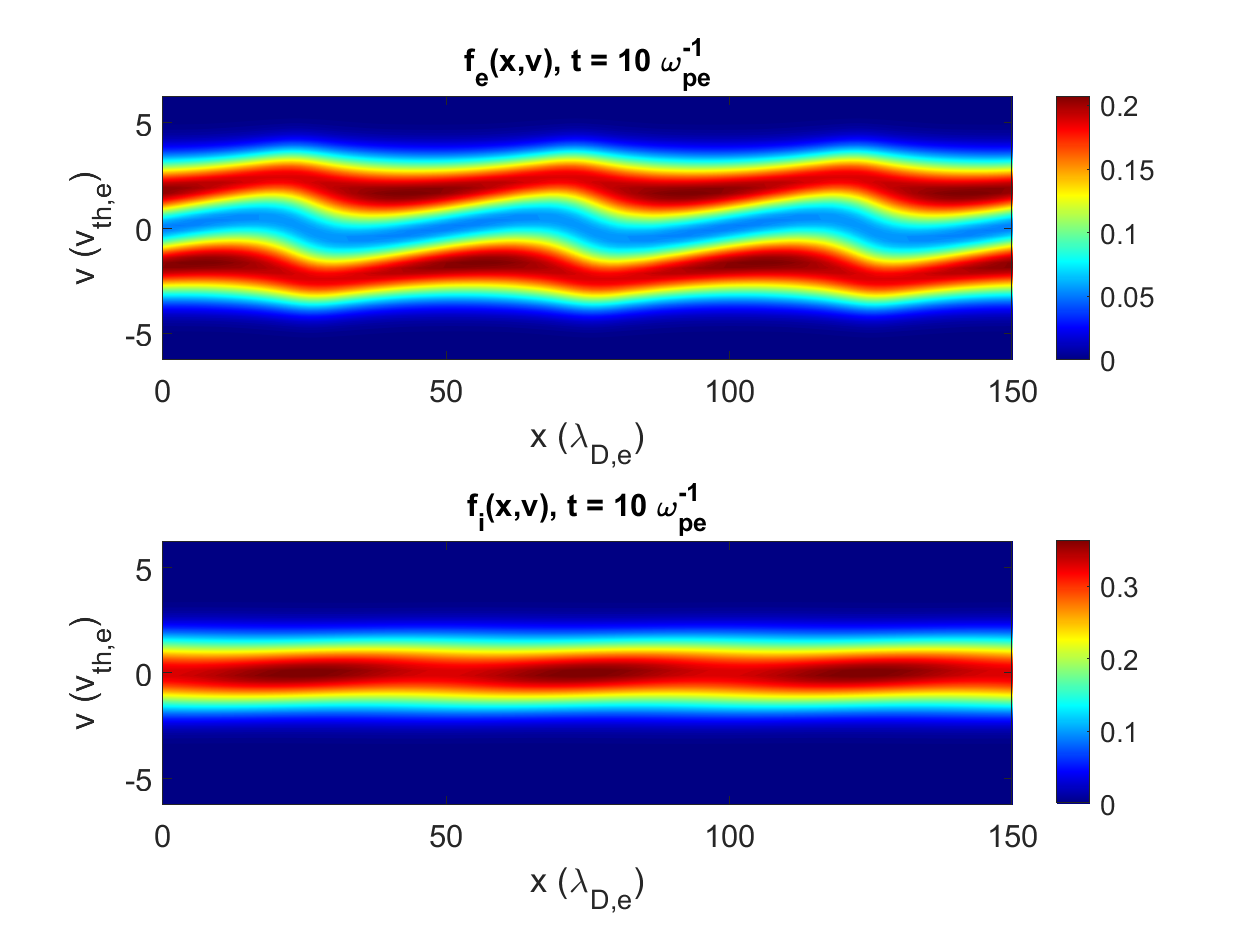}
            \includegraphics[width=0.45\linewidth]{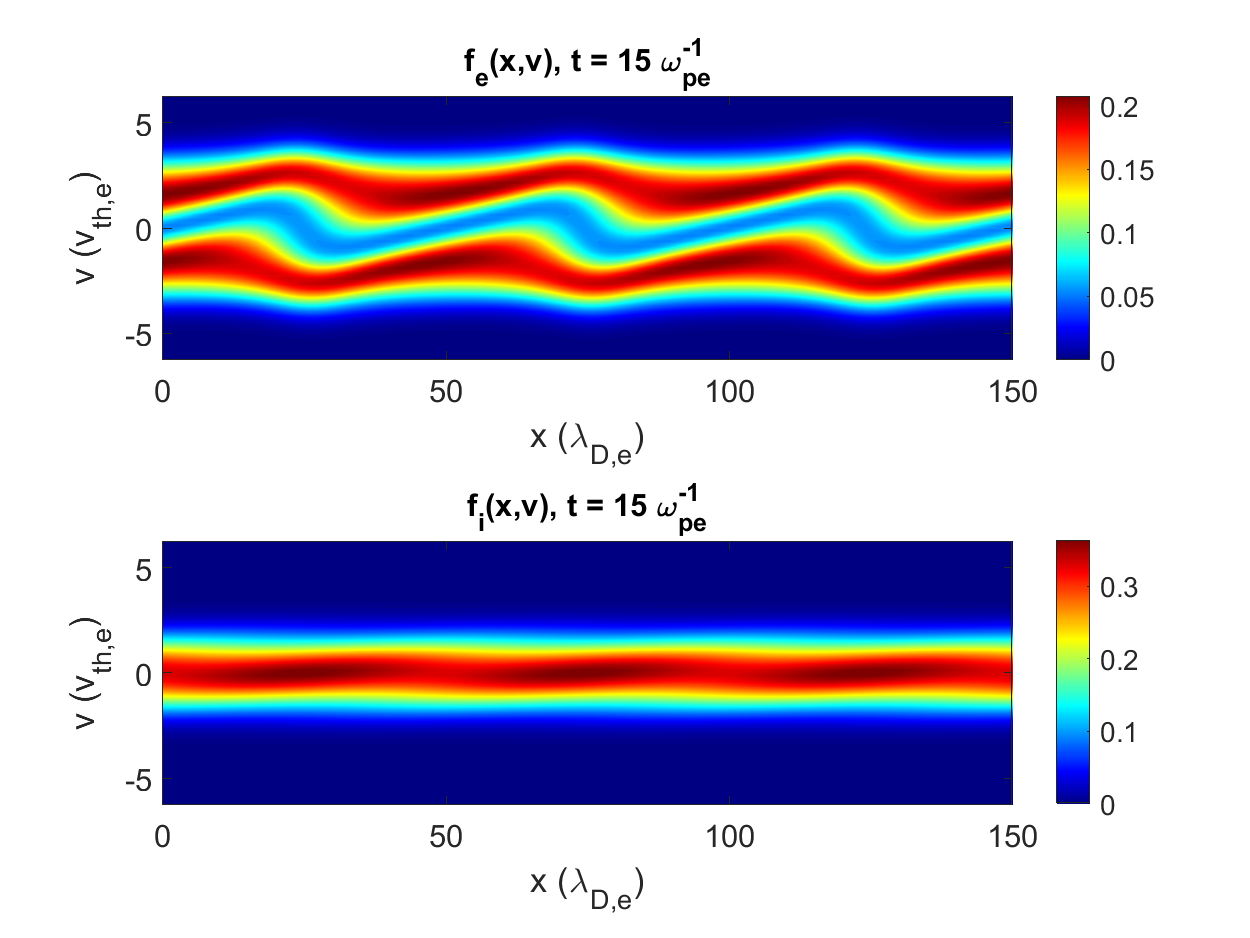}     \includegraphics[width=0.45\linewidth]{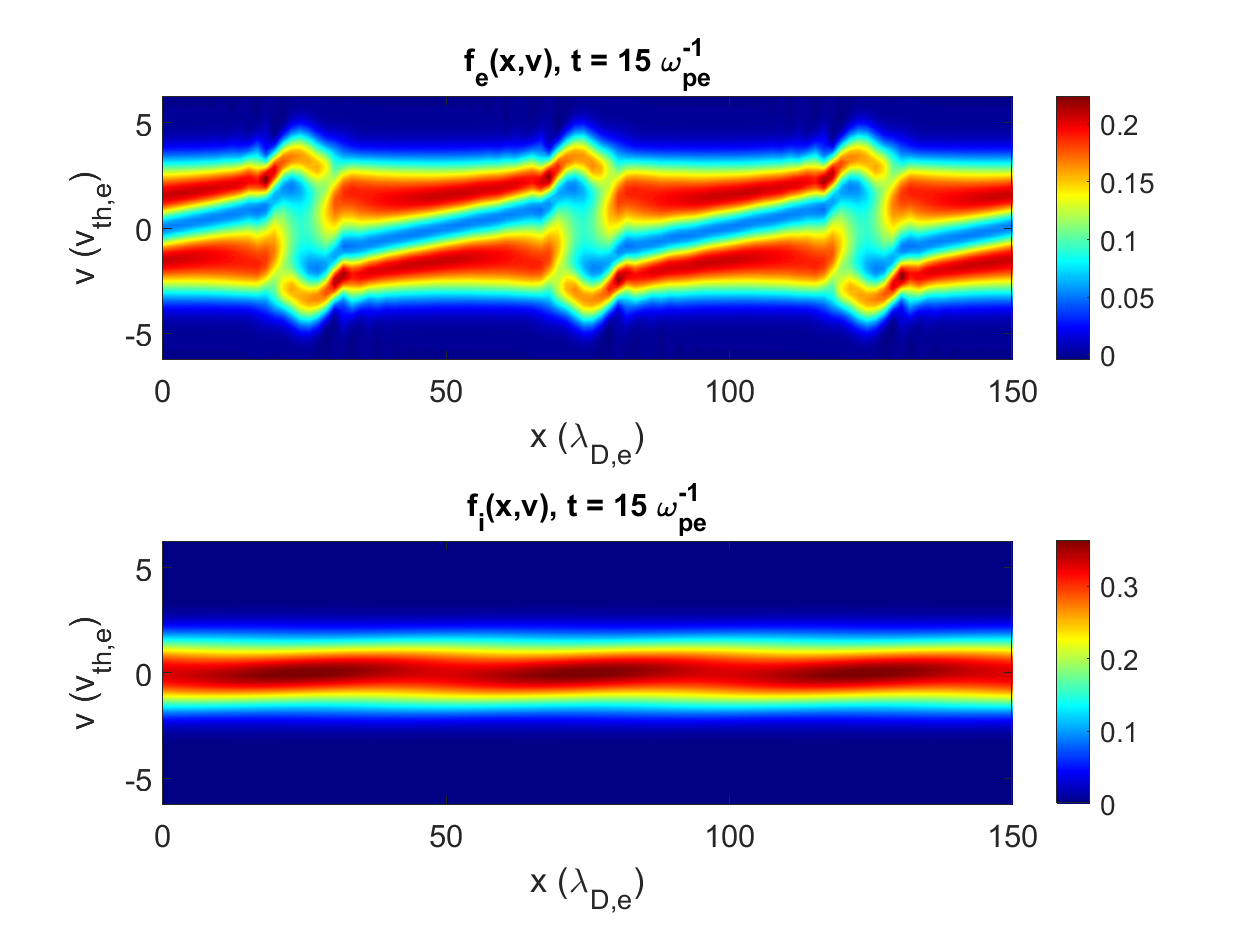}
                \includegraphics[width=0.45\linewidth]{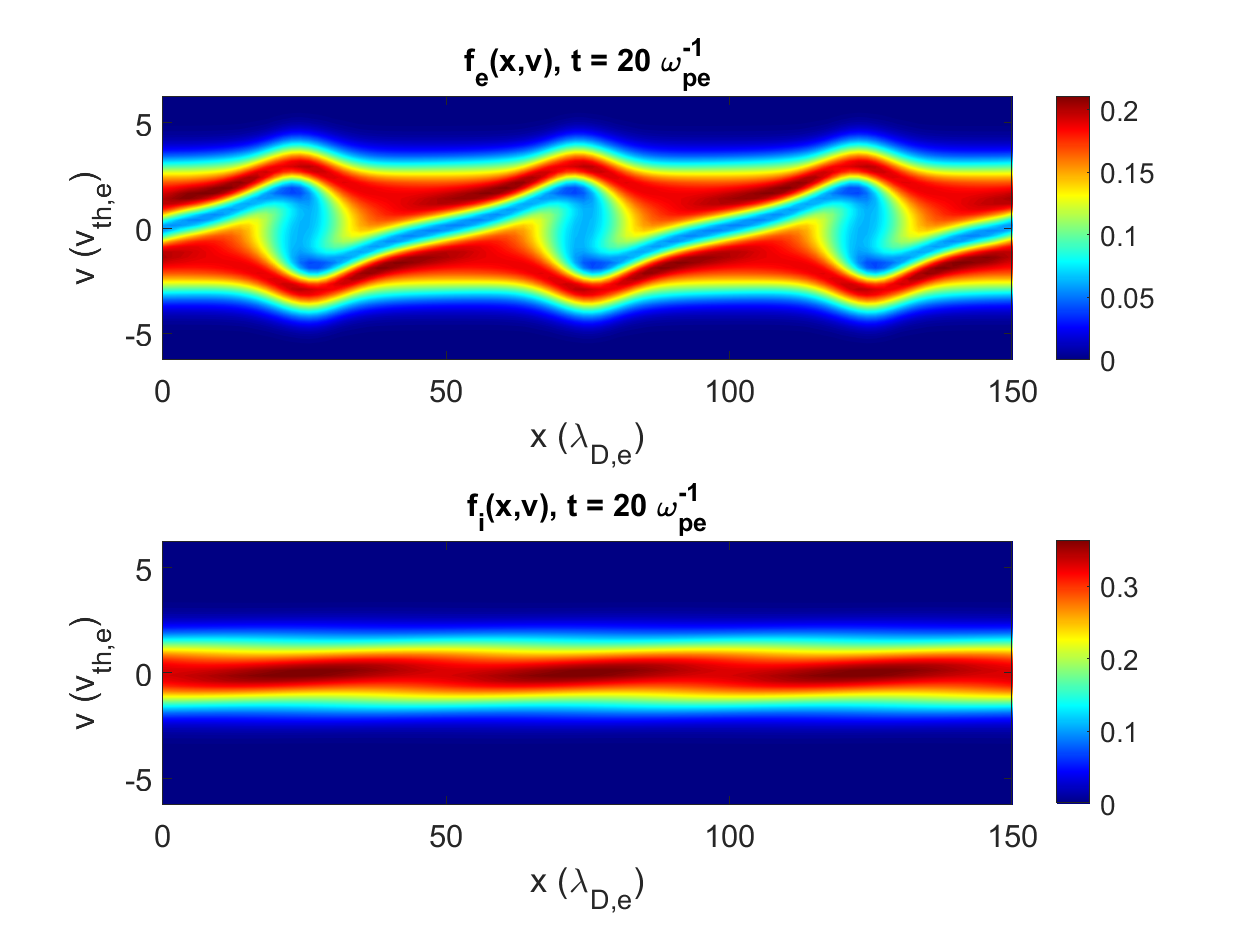}     \includegraphics[width=0.45\linewidth]{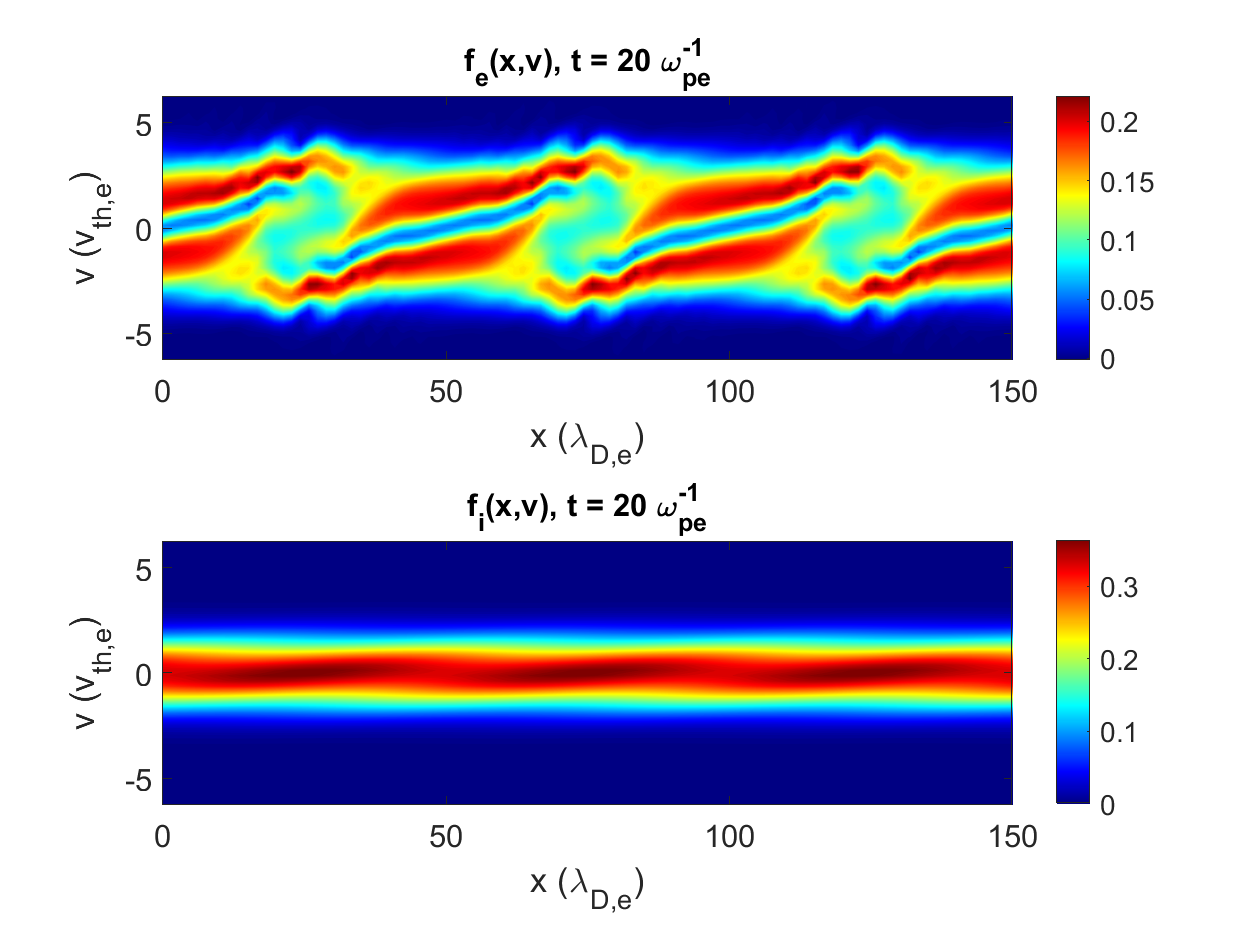}
    
 \caption[font=small,labelfont=bf]{Snapshots of the electron and ion distribution functions in phase space $(x,v)$ at different times. The left column corresponds to the VP case, and the right column to the QN case. A distinct evolution of $f_e$ is observed between the two scenarios, especially during the nonlinear phase.}
    \label{fig_2}
\end{figure}

We simulate the time evolution of $f_i$ and $f_e$ up to 20 plasma periods $\omega_{p,e}^{-1}$, while the time step is $\Delta t =10^{-4}$. In Fig.~\ref{fig_1} the initial conditions of the distribution functions $f_e$ and $f_i$ are plotted  in phase space, while in Fig.~\ref{fig_2} we provide snapshots at $t=5,10,15$ and $20$ $\omega_{p,e}^{-1}$. A distinct evolution of $f_e$ is observed between the two scenarios, particularly during the nonlinear phase. In the QN (QN) case, phase-space vortices form earlier than in the VP simulation and have different shapes. 

\begin{figure}[h!]
    \centering
    \includegraphics[width=0.45\linewidth]{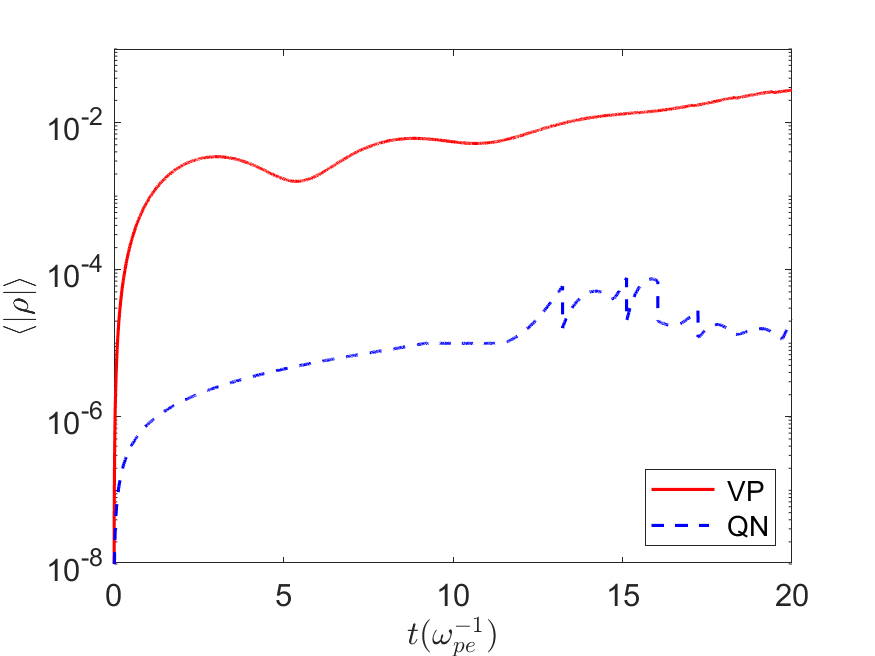}     \includegraphics[width=0.45\linewidth]{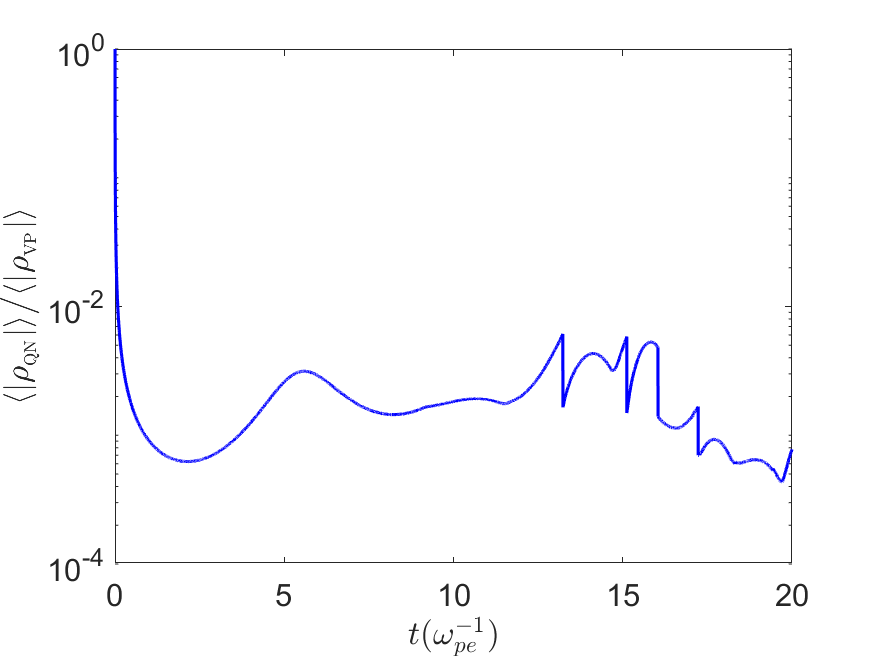}
    \caption[font=small,labelfont=bf]{Left: Evolution of the average charge density modulus $\langle|\rho|\rangle$ in the standard VP scenario (solid red line) vs the Dirac-constrained QN scenario (dashed blue line). The QN charge density, although non-zero, remains consistently three orders of magnitude smaller than in the VP case, even in the vortex saturation stage (right).}
    \label{fig_3}
\end{figure}

\begin{figure}[h!]
    \centering
    \includegraphics[width=0.45\linewidth]{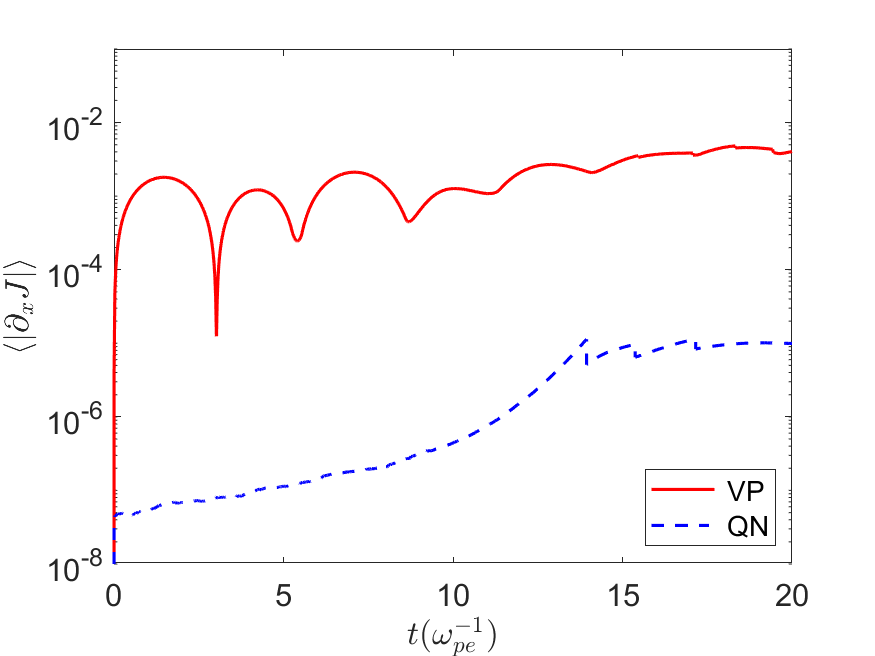}     \includegraphics[width=0.45\linewidth]{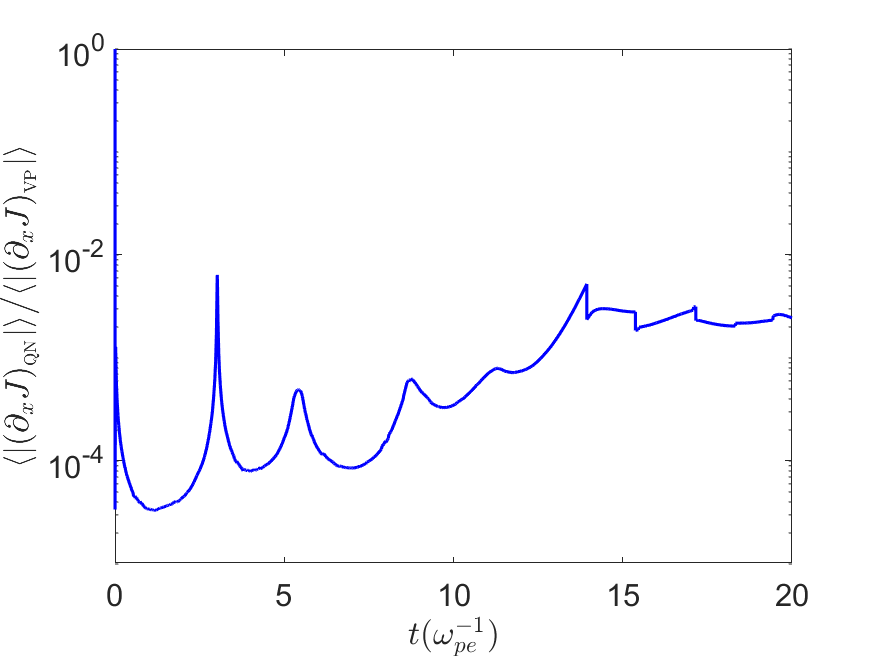}
    \caption[font=small,labelfont=bf]{Left: Evolution of  $\langle|\partial_x J|\rangle$ in the standard VP scenario (solid red line) vs the Dirac-constrained QN scenario (dashed blue line). Although non-zero, this quantity remains consistently at least three orders of magnitude smaller than in the VP case, even in the vortex saturation stage (right).}
    \label{fig_4}
\end{figure}

\begin{figure}
    \centering
    \includegraphics[width=0.45\linewidth]{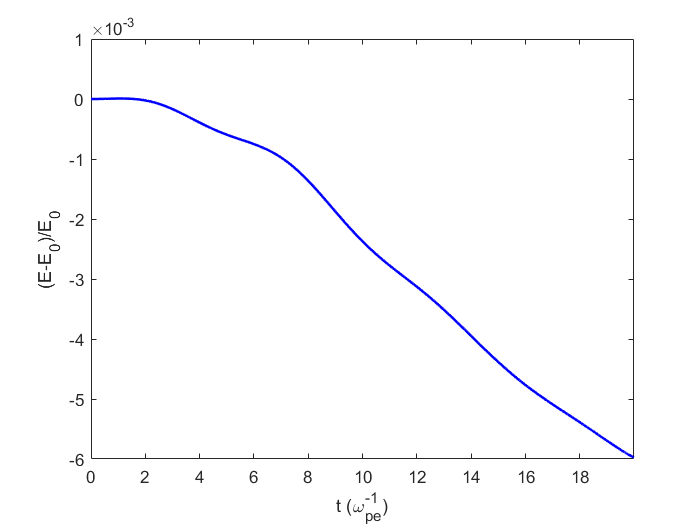}
    \includegraphics[width=0.45\linewidth]{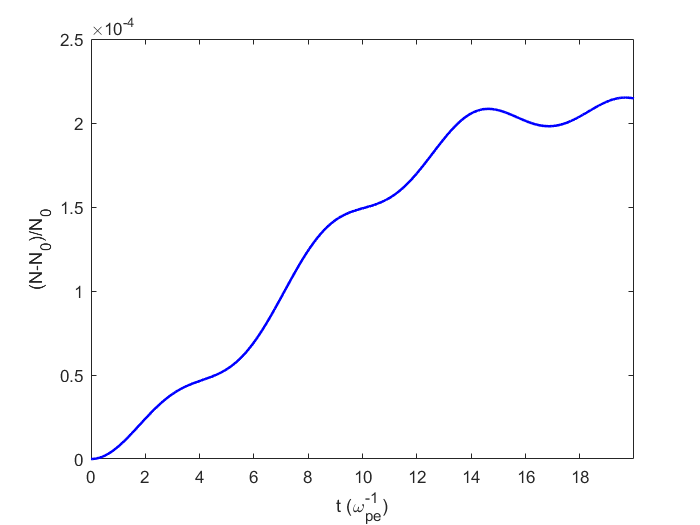}
\caption[font=small,labelfont=bf]{Relative energy error (left) and relative particle error (right) in the Vlasov-Poisson simulation. The particle number and energy are not conserved with high precision due to the non-conservative nature of the simulation algorithm, interpolation errors, and the applied filtering method.}

    \label{fig_5}
\end{figure}

In Fig.~\ref{fig_3}, we show the temporal evolution of the space-averaged modulus of the charge density, $\langle |\rho| \rangle$, as well as the ratio $\langle |\rho_{_{\mathrm{QN}}}| \rangle / \langle |\rho{_{_\mathrm{VP}}}| \rangle$, while in Fig. \ref{fig_4} we show the corresponding diagrams for the second locally conserved quantity $\bnb \cdot\bsJ=\partial_x J$. Both $\rho$ and $\partial_x J$ are three orders of magnitude smaller than their VP counterparts throughout the simulation. The initially vanishing charge density and the initial current incompressibility are not preserved to machine precision in the QN case, as the algorithm is not specifically designed to conserve energy and Casimir invariants with high accuracy. This limitation is also reflected in the relative energy and particle number errors in the Vlasov-Poisson (VP) simulation, shown in Fig.~\ref{fig_5}. These errors diminish with decreasing discretization lengths, indicating that they primarily stem from discretization and interpolation errors. VA simulations were also performed for the non-QN system yielding similar results.

\subsubsection{Electron-positron plasma ($\mu=1$)}
For the case of an electron-positron plasma, where $\mu = 1$, we assume that the distribution functions $f_e$ and $f_i$ are given by \eqref{f_e} and \eqref{f_i}, respectively, introducing a shift by a phase $\pi$ in the $x$-direction in the positron distribution, $f_i$, resulting in an initially non-vanishing charge density and a corresponding   electric field. All other simulation parameters remain unchanged, except for the time step, which is set to $\Delta t=10^{-3}$ in this case.  Thus, this setup is intended to test the conservation of the Casimir $\Phi_1$, not to mimic a physically QN regime. It ultimately demonstrates that the constrained algorithm correctly handles non-zero initial charge distributions by enforcing their preservation, in contrast to the standard VP dynamics where the charge density evolves freely. 

\begin{figure}[h!]
    \centering
    \includegraphics[width=0.45\linewidth]{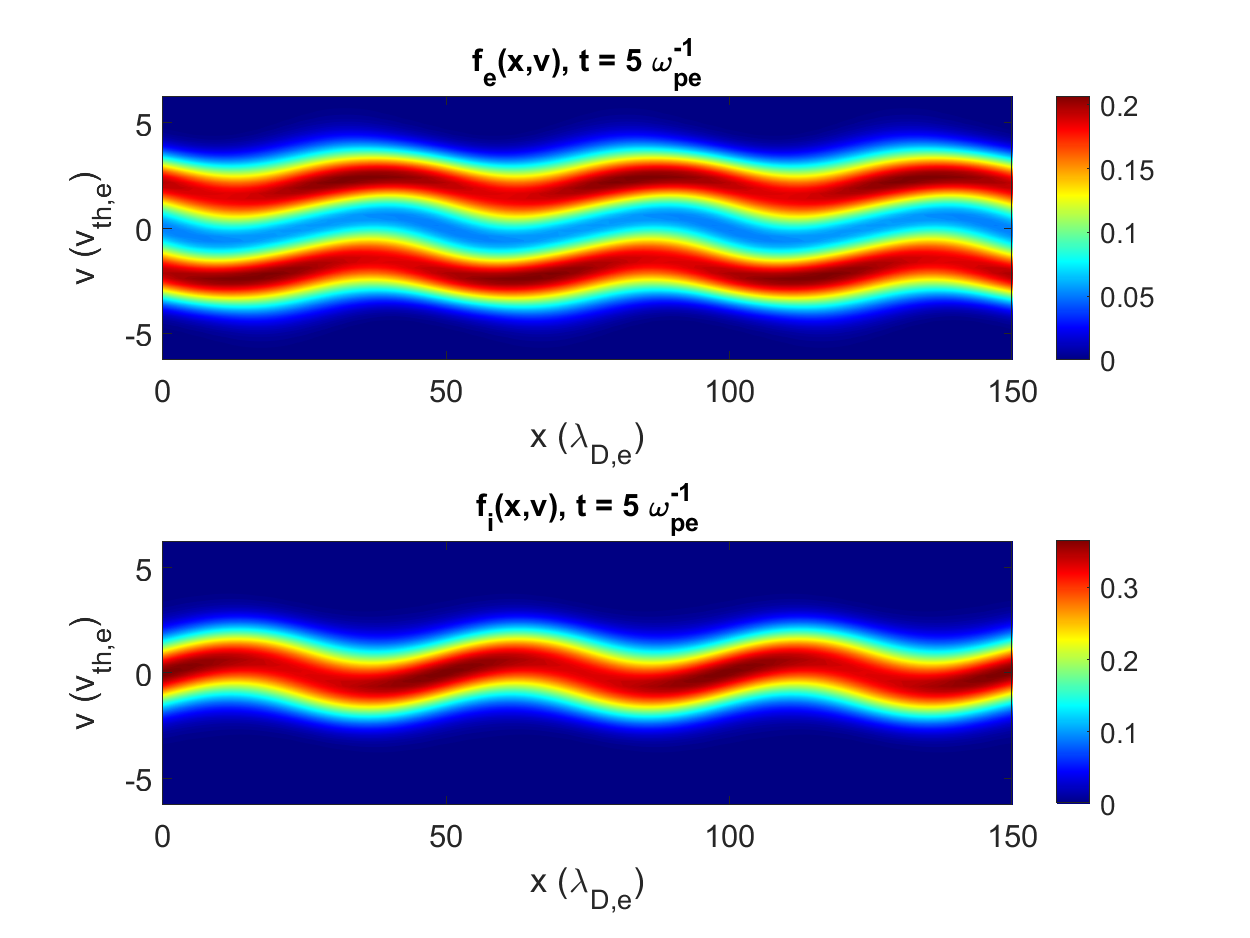}    \includegraphics[width=0.45\linewidth]{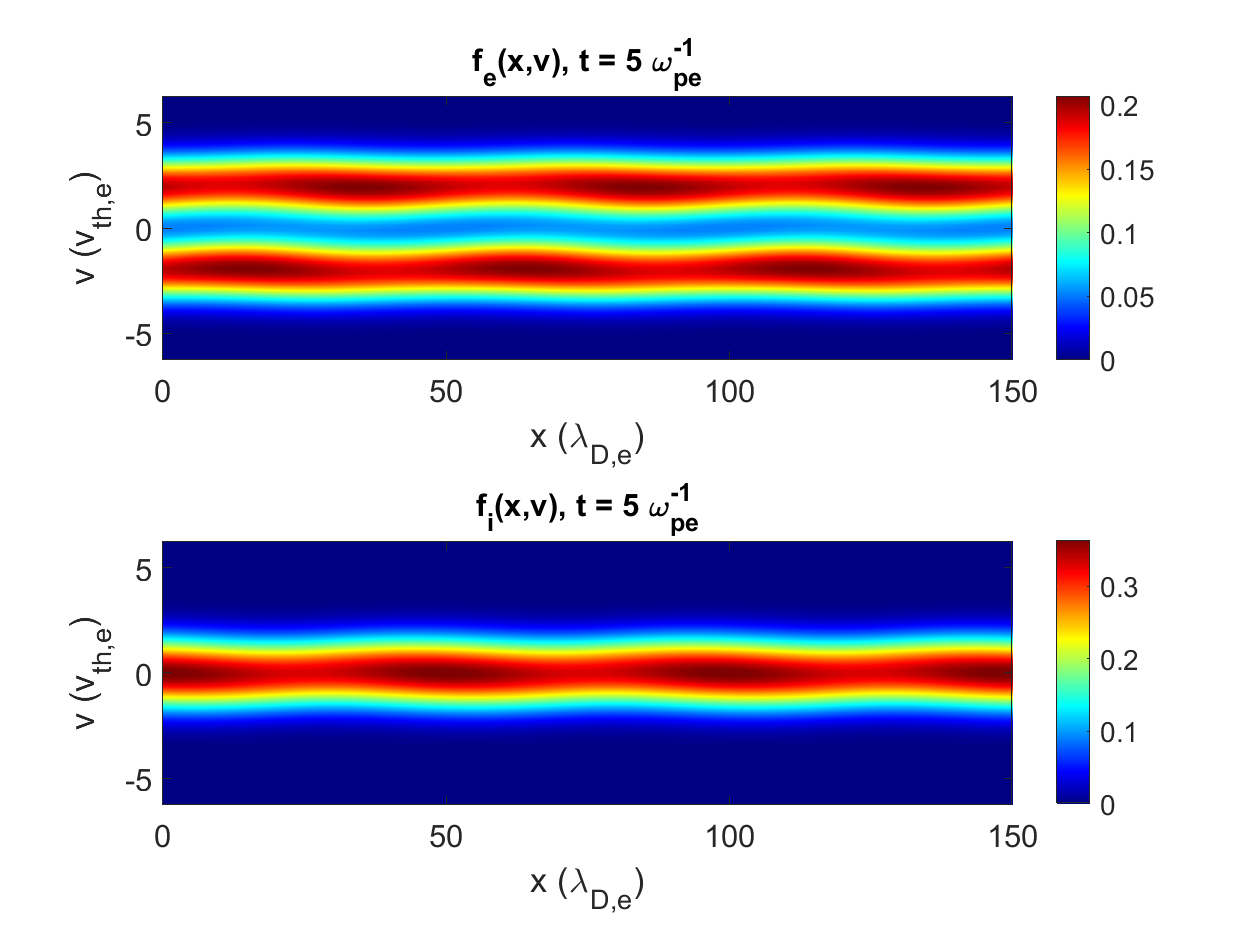}
        \includegraphics[width=0.45\linewidth]{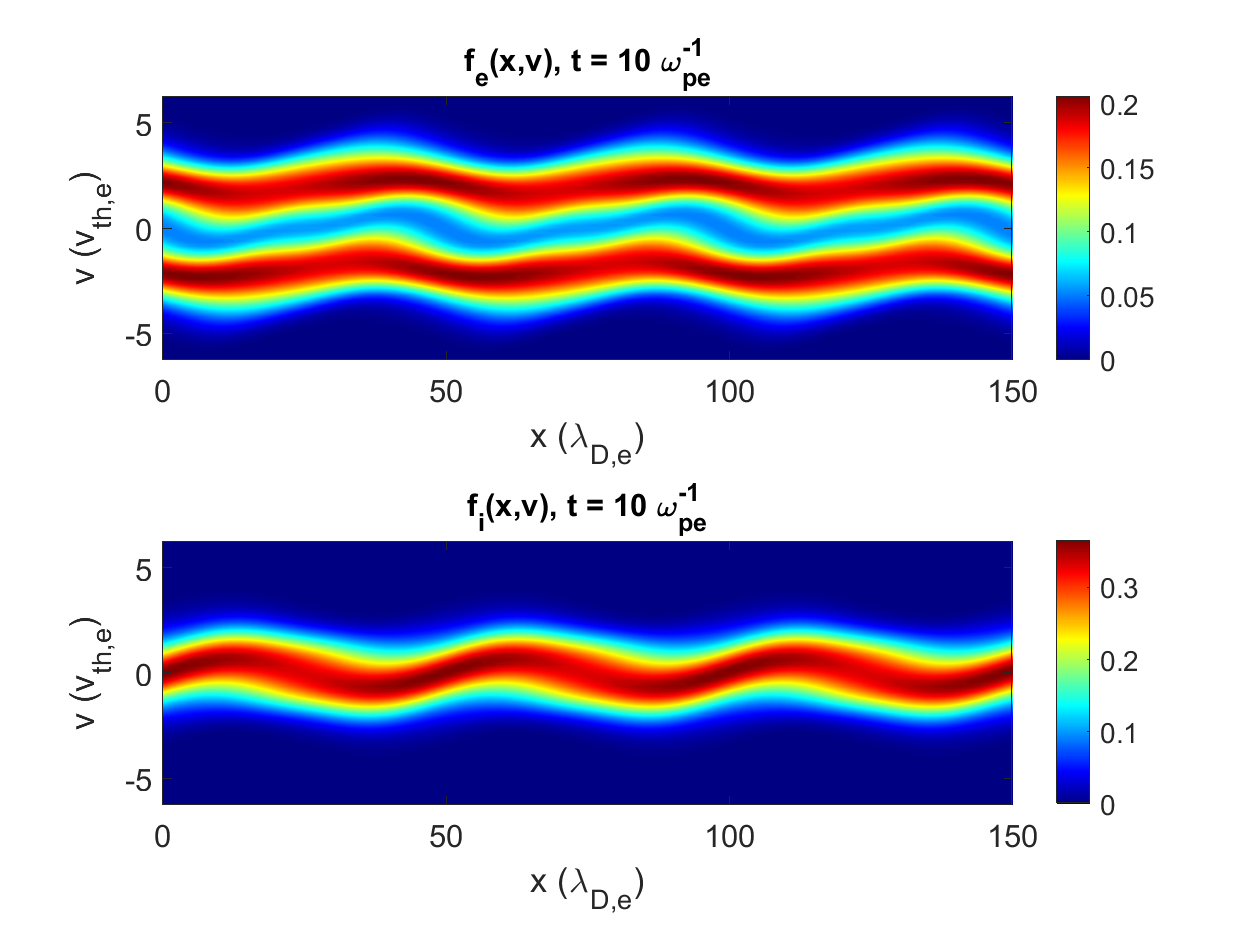}    \includegraphics[width=0.45\linewidth]{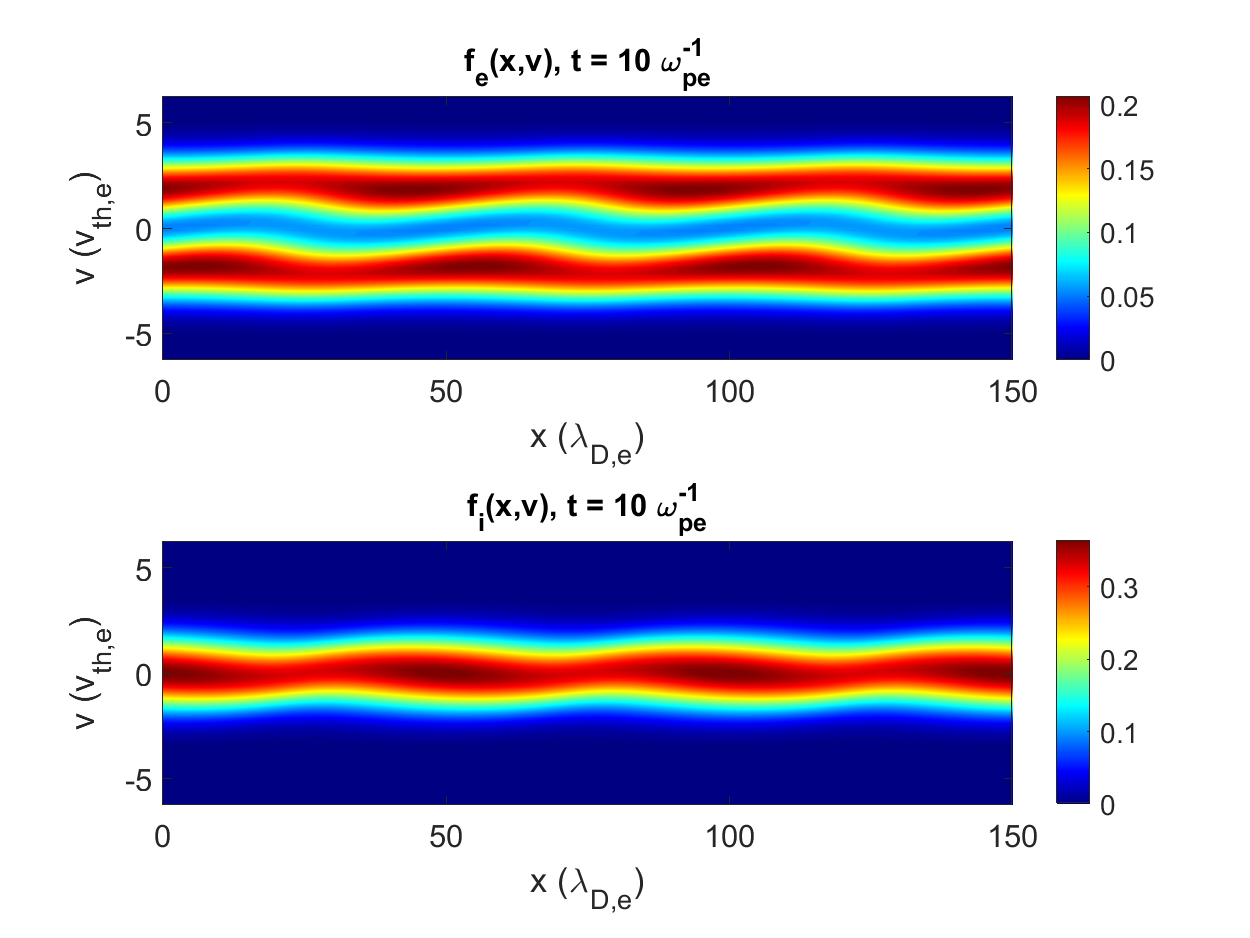}    \includegraphics[width=0.45\linewidth]{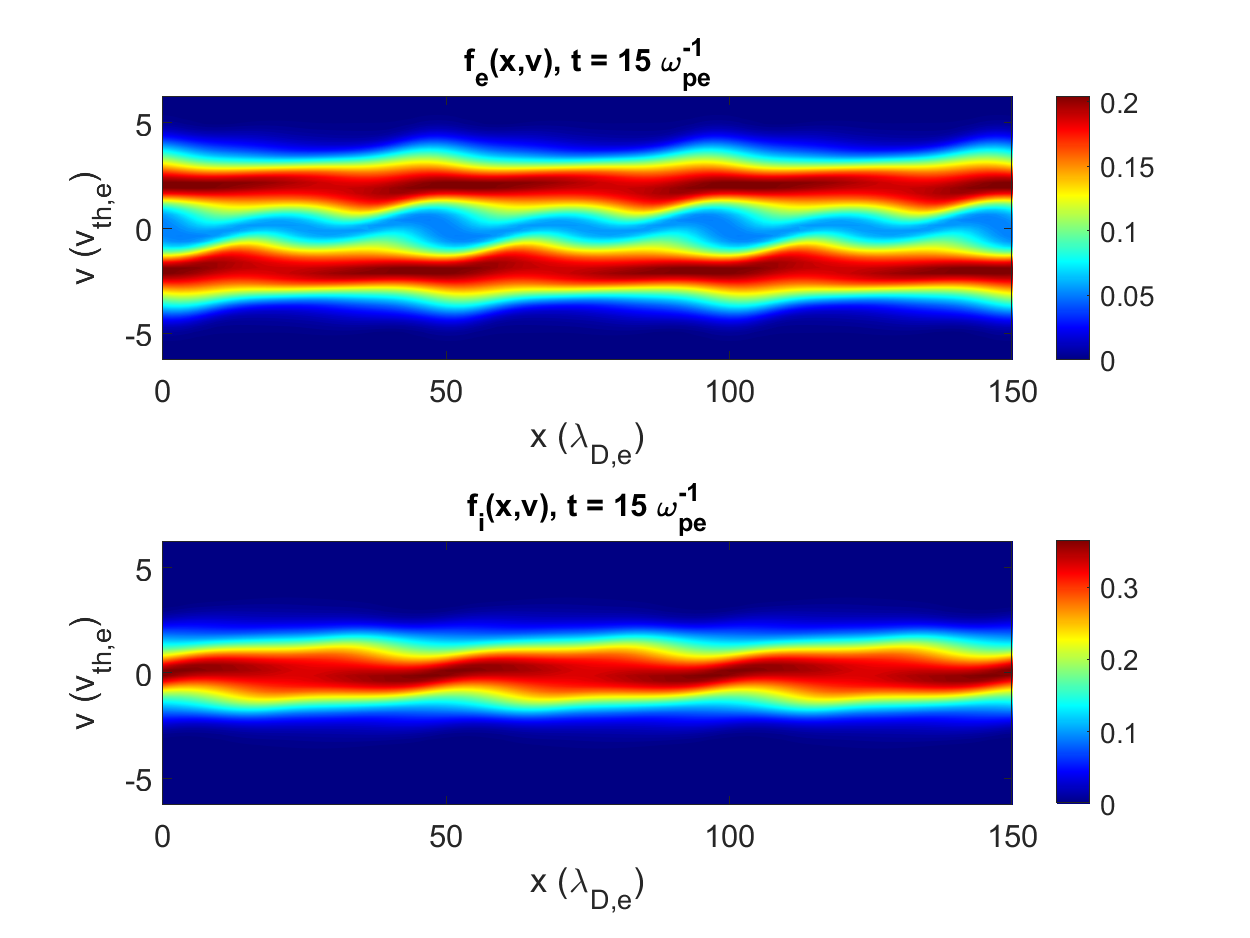}    \includegraphics[width=0.45\linewidth]{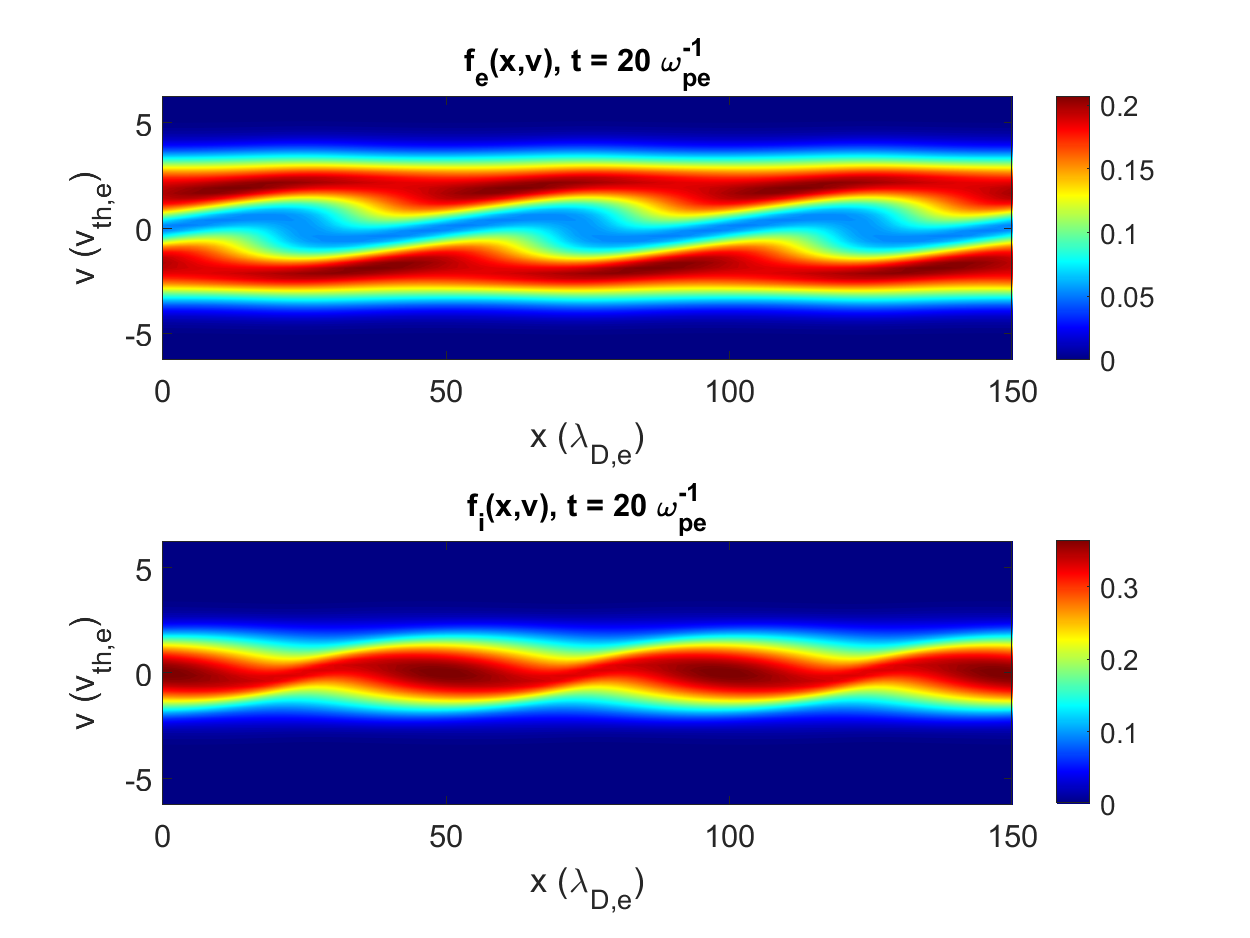}
        \includegraphics[width=0.45\linewidth]{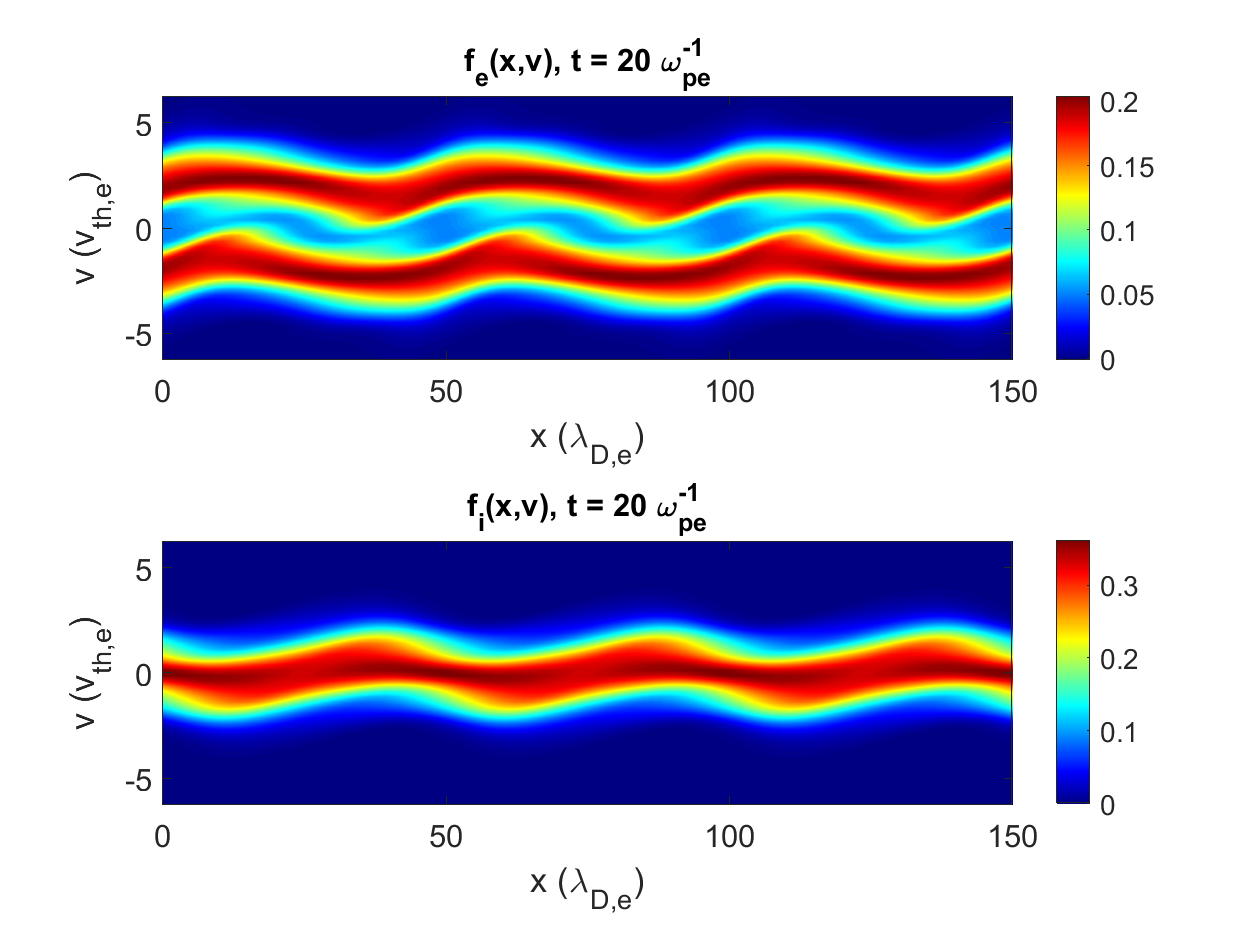}    \includegraphics[width=0.45\linewidth]{figures/fe_fi_t=20_delta=1_ep.png}
    \caption[font=small,labelfont=bf]{Snapshots of the electron and positron distribution functions in phase space $(x,v)$ at different times. The left column corresponds to the VP case, and the right column to the QN case. We observe distinct evolution for both $f_e$ and $f_i$.}
    \label{fig_6}
\end{figure}

\begin{figure}[h!]
    \centering
    \includegraphics[width=0.45\linewidth]{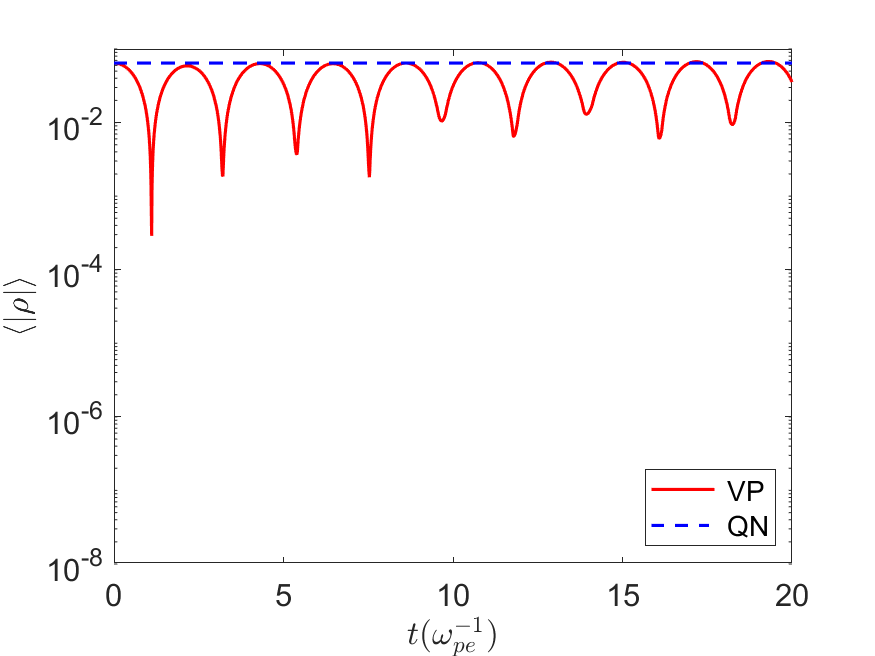}
        \includegraphics[width=0.45\linewidth]{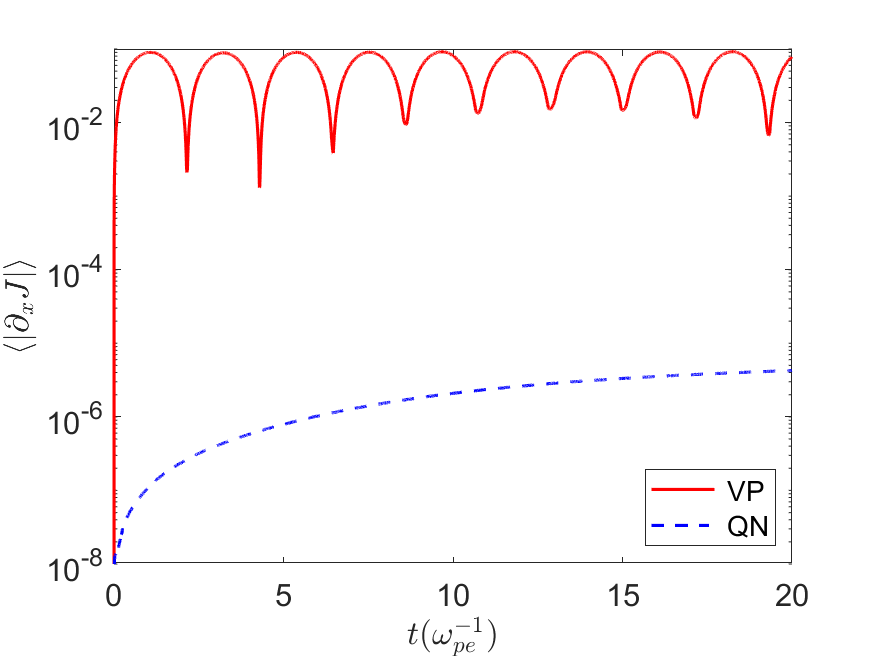}
    \caption[font=small,labelfont=bf]{Left: Evolution of the average charge density modulus $\langle|\rho|\rangle$ in the standard VP scenario (solid red line) vs the Dirac-constrained QN scenario (dashed blue line). We see that $\rho_{_{QN}}$ stays constant while $\rho_{_{VP}}$ shows significant change during the evolution. Right: The corresponding plots for the quantity $\langle|\partial_x J|\rangle$ showing that in the QN case the current incompressibility constraint is satisfied with acceptable accuracy.}
    \label{fig_7}
\end{figure}

In Fig.~\ref{fig_6}, we show contour plots of the distribution functions in phase space $(x\text{-}v)$ which reveal significant differences in the evolution of the distribution functions between the VP and the QN case. In Fig.~\ref{fig_7}, we confirm that the QN system approximately preserves the initial charge density, unlike the VP system, in which $\rho$ evolves and undergoes significant changes over time. The quantity $\partial_x J$ also exhibits significant temporal variations in the VP system, whereas in the QN case it remains approximately four orders of magnitude smaller. Thus, the current incompressibility constraint is satisfied with good precision.

 The results presented in Figs.~\ref{fig_3}, \ref{fig_4} and~\ref{fig_7} demonstrate that the numerical scheme preserves the key Casimir invariants imposed as Dirac constraints, namely the fixed charge density and the current incompressibility. This numerical preservation serves as a direct verification that the constrained dynamics has been implemented correctly. An additional verification step could, in principle, involve comparison with an analytical dispersion relation in the linear phase. However, deriving such a closed-form dispersion relation for the Dirac-constrained QN Vlasov system is significantly more involved than in the standard Vlasov–Poisson case. For this reason, a thorough dispersion-relation study, including  comparison with theory, is deferred to work focused specifically on linear validation.

\subsubsection{Estimating the significance of the Dirac forces}
To estimate the significance of the Dirac forces we consider the case of electron-proton plasma ($\mu\ll 1$) with immobile protons. To assess the relative strength of the Dirac forces acting on the electrons compared to the ``fluid forces'' due to pressure and convection, we take the zeroth and the first-order velocity moments of the electron Vlasov equation in 1D-1V to obtain:
\begin{eqnarray}
\partial_t n_e &=& -\partial_x(n_eu_e)-\xi\partial_xn_e -n_e\partial_x\xi\,, \label{0th_moment}\\
    \partial_t(n_e u_e) &=& -\partial_x P_e-\xi\partial_x(n_eu_e)-2 n_eu_e\partial_x\xi+n_e(\zeta-\eta)\,,\label{1st_moment}
\end{eqnarray}
where $u_e = n_e^{-1} \int d^3v\, f_ev$. Using \eqref{0th_moment} to reformulate \eqref{1st_moment} we find the following momentum equation:
\begin{eqnarray}
    n_e \partial_t u_e= - n_e u_e \partial_x u_e -\partial_x \tilde P_e -n_e \xi\partial_xu_e -n_eu_e \partial_x\xi +n_e(\zeta-\eta)\,, \label{mom_eq}
\end{eqnarray}
where
$$\tilde P_e = \int dv\, (v-u_e)^2f_e=P_e-n_eu_e^2\,.$$
The first two terms in the right hand side (rhs) of \eqref{mom_eq} correspond to ``fluid forces'' while the rest of the terms in the rhs are Dirac force densities. To quantify the relative strength of the Dirac forces compared to the fluid forces we investigate the evolution of the ratio:
\begin{eqnarray}
    \frac{\langle|n_e\xi\partial_xu_e +n_eu_e \partial_x\xi -n_e(\zeta-\eta)|\rangle}{\langle| n_eu_e \partial_x u_e +\partial_x \tilde P_e|\rangle}\,, \label{ratio}
\end{eqnarray}
for various characteristic lengths $L$,  from $L=25\, \lambda_{De}$ to $L=250\,\lambda_{De}$.
\begin{figure}
    \centering
    \includegraphics[width=0.45\linewidth]{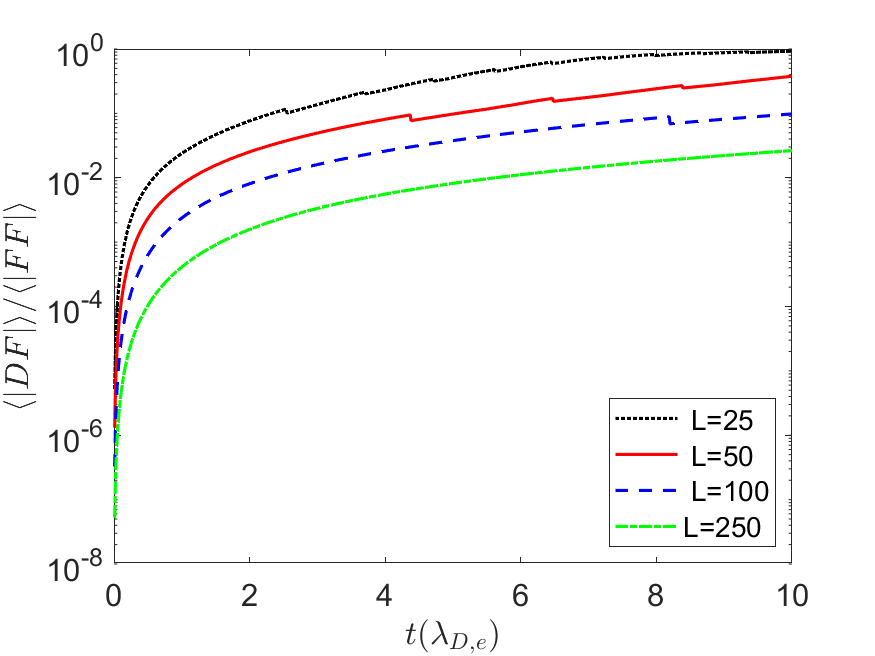}\includegraphics[width=0.45\linewidth]{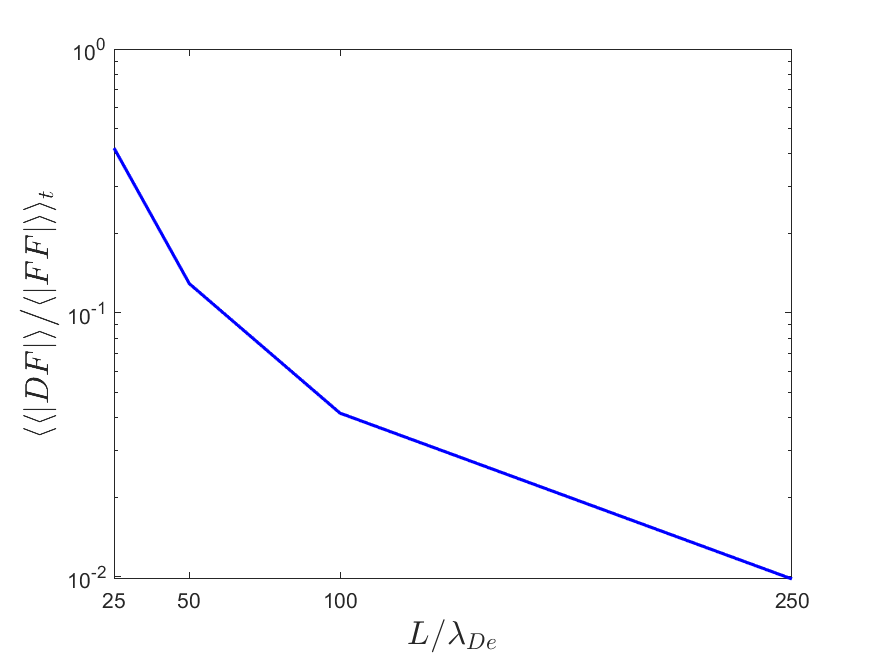}
    \caption{Left: the evolution of the ratio \eqref{ratio} of the Dirac forces over the fluid forces for a simulation with steady ion distribution. In all four cases $L=25$, $L=50$, $L=100$ and $L=250$, this ratio increases with time and becomes progressively smaller for larger length scales. Right: the time-averaged ratio versus the length scale $L$.}
    \label{fig_8}
\end{figure}
For this comparison we consider a steady ion particle density $n_i=1-\epsilon cos(2\pi k x/L)$ and the electrons are initially described by \eqref{f_e} with $V_e=0$, in order to avoid the formation of phase-space vortices and the associated fine structures which are sources of instability and limit the simulation time window especially for small scales $L$. The simulation ends at $t=10\,\omega_{pe}$ for which the quasineutrality condition is preserved with good accuracy. For later simulation times instability kicks in and quasineutrality is violated. In Fig. \ref{fig_8} we present the evolution of the ratio \eqref{ratio} for the various length scales $L$ and the time-averaged ratio. We observe that in all four cases $L=25$, $L=50$, $L=100$ and $L=250$, the ratio of the Dirac forces over the fluid forces increases with time and becomes progressively smaller for larger length scales. The time averaged ratio is over $10^{-1}$ for length scales $L<50$ meaning that the Dirac forces responsible for imposing quasineutrality are significant and thus quasineutrality is not a good approximation while it becomes a better approximation for $L>250$ where the averaged ratio is smaller than $10^{-2}$.  This comparison verifies a consistent QN scaling: Dirac forces become negligible at large system size, in agreement with the known asymptotic QN limit.

\section{Conclusions}
\label{sec_V}
In this work, we have reformulated the Vlasov-Poisson and Vlasov-Ampère systems by imposing quasineutrality as a Dirac constraint. The resulting constrained dynamical equations were derived within the noncanonical Hamiltonian framework, using the standard Hamiltonian functionals of the two models and Dirac brackets constructed via Dirac’s algorithm. In this formulation, the electric field is eliminated, and new advection terms emerge involving generalized gradient velocity and force fields. These terms enforce the quasineutrality condition (or more generally charge density conservation) and their calculation,  require, in general, solving three elliptic partial differential equations. In the 1D-1V case, however, the gradient fields can be computed explicitly.

We also performed numerical simulations using the constrained QN (QN) equations and demonstrated that the evolution of the distribution function differs significantly from that of the standard, unconstrained VP and VA systems. We observed that in the QN simulation, the charge density remains consistently over two orders of magnitude smaller than that in the VP simulation, validating the effectiveness of the method used to impose quasineutrality. The fact that the charge density does not vanish to machine precision is attributed to the numerical scheme, which is not specifically designed to conserve invariants; however, the accuracy improves with finer grids and smaller time steps. Finally, we performed a comparative analysis of the Dirac forces relative to the fluid forces, showing that the former becomes less significant at larger length scales. These numerical experiments serve as verification tests for the new model, since the observed preservation of charge density and current incompressibility, demonstrate that the discretized implementation satisfies the Dirac-imposed constraints.

A detailed analytical dispersion-relation analysis, enabling direct comparisons in the linear regime, would also be a valuable verification test of the present implementation. However, such an analysis lies beyond the scope of this introductory study and is deferred to future work. In addition, the extension of the Dirac-constraint framework to the full Vlasov–Maxwell system, allowing for self-consistent magnetic-field generation, will be presented in a separate forthcoming paper. Future research will also investigate Casimir- or, more generally, structure-preserving discretizations, as well as applications of the QN Dirac-constraint methodology to hybrid fluid–kinetic models such as those proposed in~\cite{Kaltsas2021}.

\section*{Acknowledgments}
The contributions of DAK and GNT were carried out within the framework of the participation of the University of Ioannina in the National Programme for Controlled Thermonuclear Fusion of the Hellenic Republic. JWB was supported by the U.S. Department of Energy, Office of Science, Office of Advanced Scientific Computing Research, as a part of the Mathematical Multifaceted Integrated Capability Centers program, under Award Number DE-SC0023164.  Both JWB and PJM received support from the DOE Office of Fusion Energy Sciences under DE-FG02-04ER-54742.  PJM  also acknowledges support via NSF \# DMS-1928930 during his stay at the Simons Laufer Mathematical Sciences Institute in the Fall of 2025.   E. Tassi acknowledges support from the GNFM.  

\section*{Conflict of interest}
The authors have no conflict of interest.


 \begin{appendices}

\section{Jacobi identity for the VA bracket}
\label{app_1}
Following the analysis presented in the Appendix of \cite{Morrison2013}, we find that 
\begin{eqnarray}
    \{\{F,G\}_{_{VA}},H\}_{_{VA}}+cyc
    = \sum_s\frac{q_s}{\epsilon_0 m_s^2}\int  d^3xd^3v\, \left(\bnb\times \frac{\delta H}{\delta \bsE}\right)\cdot\left(\bnb_v\frac{\delta F}{\delta f_s}\times \bnb_v\frac{\delta G}{\delta f_s} \right)+cyc\,, \label{jacobiator_1}
\end{eqnarray}
where $F,G,H$ are three arbitrary functionals on the functional phase space and $cyc$ denotes their cyclic permutation. Jacobi identity is satisfied if the right hand side of \eqref{jacobiator_1} is zero, which is not generally true.

However, if the electric field is irrotational and thus $\bsE=-\bnb\phi$, then the functional derivative of an arbitrary functional $F$ with respect to $\bsE$, is connected with the functional derivative $\delta F/\delta\phi$ as follows:
\begin{eqnarray}
    \frac{\delta F}{\delta\bsE} = - \bnb\Delta^{-1}\frac{\delta F}{\delta \phi}\,, \label{deltaE_deltaPhi}
\end{eqnarray}
where $\Delta^{-1}$ denotes the inverse Laplacian operator, i.e., the solution operator to $\Delta f=g$, that can be defined via convolution with the Green’s function as in Sec.~\ref{sec_II} with appropriate boundary conditions. In view of \eqref{deltaE_deltaPhi}, the right hand side of \eqref{jacobiator_1} is trivially zero and thus the Jacobi identity is satisfied, making bracket \eqref{Poisson_bracket_VA} a Poisson bracket. 

To prove \eqref{deltaE_deltaPhi} we start by viewing $F$ as a functional of $\bsE$ and then as a functional of $\phi(\bsE)$. While the respective expressions may differ in form, they represent the same functional, so we write
$F(f_s,\bsE)=\bar{F}(f_s,\phi)\,.$
The variations of $F$ and $\bar{F}$ with respect to  their respective field variables must therefore agree:
\begin{eqnarray}
    \int d^3x \frac{\delta F}{\delta\bsE}\cdot\delta\bsE=\int d^3x \frac{\delta \bar{F}}{\delta\phi}\delta\phi\,.\label{deltaF_deltaFbar_1}
\end{eqnarray}
Since $\bsE=-\bnb\phi$, then $\phi = - \Delta^{-1}\bnb\cdot\bsE$ and therefore Eq.~\eqref{deltaF_deltaFbar_1} becomes
\begin{eqnarray}
       \int d^3x \frac{\delta F}{\delta\bsE}\cdot\delta\bsE=-\int d^3x \frac{\delta \bar{F}}{\delta\phi} \Delta^{-1}\bnb\cdot\delta\bsE\,.\label{deltaF_deltaFbar_2} 
\end{eqnarray}
The variation $\delta \bsE$ is thus not arbitrary but constrained to lie in the space of irrotational fields. Using the self-adjointness of $\Delta^{-1}$ and integrating by parts we find:
\begin{eqnarray}
       \int d^3x \frac{\delta F}{\delta\bsE}\cdot\delta\bsE=-\int d^3x \,\delta\bsE\cdot \bnb \Delta^{-1}\frac{\delta \bar{F}}{\delta\phi} \,.\label{deltaF_deltaFbar_2} 
\end{eqnarray}
Since this equation must hold for all admissible variations $\delta\bsE$, it follows that \eqref{deltaE_deltaPhi} must hold.

 \section{Calculation of the C-matrix elements}
 \label{app_2}
Here, we present the procedure for calculating the elements of the constraint matrix $C$, namely $C_{ij}$, in the VP case, where the Poisson bracket is given by Eq.~\eqref{Poisson_bracket_VP}. For these calculations we need the functional derivatives of the constraints $\Phi_1$ and $\Phi_2$ as given by \eqref{Phi_1_2} and \eqref{Phi_2}, respectively:
\begin{eqnarray}
    \frac{\delta \Phi_1}{\delta f_i} = \delta(\bsx-\bsx')\,, \quad \frac{\delta \Phi_1}{\delta f_e} = -\delta(\bsx-\bsx')\,,\nn\\
    \frac{\delta \Phi_2}{\delta f_i} = -\bsv\cdot\bnb' \delta(\bsx-\bsx')\,, \quad \frac{\delta\Phi_2}{\delta f_e} = \bsv \cdot\bnb'\delta (\bsx-\bsx')\,.
\end{eqnarray}
It is straightforward to see that the entry \(C_{11}(\bsx,\bsx')=\{\Phi_1(\bsx),\Phi_1(\bsx')\}\) is zero. For \(C_{12}(\bsx,\bsx')\) we have
 \begin{eqnarray}
 C_{12}(\bsx,\bsx')&=&\{\Phi_1(\bsx),\Phi_2(\bsx')\}-\int\int d^3x''d^3v\,\left(\frac{f_i}{m_i}+\frac{f_e}{m_e}\right)[\delta(\bsx-\bsx''),\bsv\cdot\nb''\delta(\bsx',\bsx'')]_{x'',v}\nn\\
 &&=-\int\int d^3x''d^3v\,\left(\frac{f_i}{m_i}+\frac{f_e}{m_e}\right)\nb''\delta(\bsx-\bsx'')\cdot\nb''\delta(\bsx'-\bsx'')\nn\\
&& =\int d^3v \nb\cdot\left[\left(\frac{f_i}{m_i}+\frac{f_e}{m_e}\right)\nb\delta(\bsx'-\bsx)\right]=\Lc\delta(\bsx'-\bsx)\,.
\end{eqnarray}  
Similarly we can find that: 
\begin{eqnarray}
C_{21}(\bsx,\bsx')=-\Lc'\delta(\bsx-\bsx')\,.
\end{eqnarray}
For the calculation of $C_{22}(\bsx,\bsx')$ we consider the bracket $\{\Phi_2(\bsx),\Phi_2(\bsx')\}$: 
\begin{eqnarray}
C_{22}(\bsx,\bsx')&=&\{\Phi_{2}(\bsx),\Phi_2(\bsx')\}\nn\\
&=&\int\int d^3x''d^3v \bigg\{\left(\frac{f_i}{m_i}+\frac{f_e}{m_e}\right)[\bsv\cdot\nb''\delta(\bsx-\bsx''),\bsv\cdot\nb''\delta(\bsx'-\bsx'')]_{x'',v}\,.
\end{eqnarray}
To proceed, let us compute separately  the particle bracket:
\begin{eqnarray}
&[\bsv\cdot\bnb''\delta(\bsx-\bsx''),\bsv\cdot\bnb''\delta(\bsx'-\bsx'')]_{x'',v}\nn\\
&= \partial_j''[v_k\partial_k''\delta(\bsx-\bsx'')]\partial_{v_j}[v_\ell\partial_\ell''\delta(\bsx'-\bsx'')]
-\partial_j''[v_k\partial_k''\delta(\bsx'-\bsx'')]\partial_{v_j}[v_\ell\partial_\ell''\delta(\bsx-\bsx'')]\nn\\
&=v_k\partial_k''\partial_j''\delta(\bsx-\bsx'')\partial_j''\delta(\bsx'-\bsx'')-v_k\partial_k''\partial_j''\delta(\bsx'-\bsx'')\partial_j''\delta(\bsx-\bsx'') \nn\\
&=(\bsv\cdot \bnb'') \bnb'' \delta(\bsx-\bsx'') \cdot \bnb'' \delta(\bsx'-\bsx'') - (\bsv\cdot \bnb'') \bnb'' \delta(\bsx'-\bsx'') \cdot \bnb'' \delta(\bsx-\bsx'')\,.
\end{eqnarray}
Therefore, $C_{22}$ becomes:
\begin{eqnarray}
C_{22}(\bsx,\bsx')&=&\int\int d^3x''d^3v \bigg\{\left(\frac{f_i}{m_i}+\frac{f_e}{m_e}\right)[\bnb''\delta(\bsx'-\bsx'')\cdot (\bsv\cdot\bnb'')\bnb''\delta(\bsx-\bsx'')\nn\\
&&-\bnb''\delta(\bsx-\bsx'')\cdot (\bsv\cdot\bnb'')\bnb''\delta(\bsx'-\bsx'')]\,.
\end{eqnarray}
Integrating by parts and neglecting boundary terms we find:
\begin{eqnarray}
C_{22}(\bsx,\bsx')=\int \int d^3x d^3v \bigg\{\delta(\bsx-\bsx'')\bnb''\cdot\left[\left(\frac{f_i}{m_i}+\frac{f_e}{m_e}\right)(\bsv\cdot\bnb'')\bnb''\delta(\bsx'-\bsx'')\right]\nn\\
+\delta(\bsx-\bsx'')(\bsv\cdot\bnb'')\bnb''\cdot\left[\left(\frac{f_i}{m_i}+\frac{f_e}{m_e}\right)\bnb''\delta(\bsx'-\bsx'')\right]\bigg\}\nn\\
=\bnb\cdot\int d^3v \bigg\{\left(\frac{f_i}{m_i}+\frac{f_e}{m_e}\right)(\bsv\cdot\bnb) \bnb\delta(\bsx'-\bsx)
+\bnb\delta(\bsx'-\bsx)\cdot\bnb\left[\bsv\left(\frac{f_i}{m_i}+\frac{f_e}{m_e}\right)\right]\bigg\}\nn\\
=\bnb\cdot\left[(\bsM\cdot\bnb)\bnb \delta(\bsx'-\bsx)+\bsM \Delta \delta(\bsx'-\bsx)+\bnb\delta(\bsx'-\bsx)\cdot\bnb\bsM\right]\,,
\end{eqnarray}
where $\bsM$ is given by \eqref{M}.

%

\appendix

\medskip

\end{appendices}

\printbibliography

\end{document}